\documentclass[a4paper,11pt]{article}
\pdfoutput=1
\usepackage{jheppub}

\newcommand{\beq}{\begin{eqnarray}}
\usepackage{subfigure}

\newcommand{\eeq}{\end{eqnarray}}
\newcommand{\bea}{\begin{eqnarray}}
\newcommand{\eea}{\end{eqnarray}}
\newcommand{\eq}[1]{Eq.~(\ref{#1})}

\newcommand{\U}{\text{U}}

\newcommand{\mti}{\tilde{M}}

\newcommand{\br}{\textrm{BR}}
\newcommand{\tev}{\,\textrm{TeV}}
\newcommand{\gev}{\,\textrm{GeV}}
\newcommand{\mev}{\,\textrm{MeV}}
\newcommand{\ev}{\,\textrm{eV}}
\newcommand{\st}{\sin^2\theta}
\newcommand{\cost}{c_{\theta}}
\newcommand{\sint}{s_{\theta}}
\newcommand{\mphi}{m_{\phi}}

\newcommand{\mpl}{M_{\rm{Pl} } }
\newcommand{\neff}{N_{\rm{eff} }}
\newcommand{\lbr}{\Lambda_{{\rm br} }}
\newcommand{\lbrmax}{\Lambda_{\rm br }^{\rm{max} } }

\preprint{CTPU-16-25}

\title{Phenomenology of relaxion-Higgs mixing}

\author[a,b]{Thomas Flacke,}
\author[c]{Claudia~Frugiuele,}
\author[c]{Elina~Fuchs,}
\author[c]{Rick~S.~Gupta}
\author[c]{and Gilad~Perez}

\affiliation[a]{Center for Theoretical Physics of the Universe, Institute for Basic Science (IBS),\\
Daejeon, 34051, Korea}
\affiliation[b]{Department of Physics, Korea University,\\
Seoul 136-713, Korea}
\affiliation[c]{Department of Particle Physics and Astrophysics,\\
Weizmann Institute of Science, Rehovot 76100, Israel}

\emailAdd{flacke@ibs.re.kr}
\emailAdd{claudia.frugiuele@weizmann.ac.il}
\emailAdd{elina.fuchs@weizmann.ac.il}
\emailAdd{rsgupta@weizmann.ac.il}
\emailAdd{gilad.perez@weizmann.ac.il}

\abstract{We show that the  relaxion  generically stops its rolling at a point that  breaks CP leading to relaxion-Higgs mixing.
This opens the door to a variety of observational probes since the possible relaxion mass spans a broad range from sub-eV to the GeV scale.
 We derive constraints from current experiments (fifth force, astrophysical and cosmological probes, beam dump, flavour, LEP and LHC) and  present  projections from future experiments such as NA62, SHiP and PIXIE. We find that a large region of the parameter space is already under the experimental scrutiny. All the experimental constraints we derive are equally applicable for general Higgs portal models.    In addition, we show that simple multiaxion (clockwork) UV completions suffer from a mild fine tuning problem, which increases with the number of sites.  These results favour a cut-off scale lower than the existing theoretical bounds.
}

\begin{document}
\maketitle\flushbottom

\section{Introduction}
\label{intro}

Relaxion models offer a new perspective on the hierarchy problem~\cite{kaplan}.
The weak scale is obtained in a dynamical way as the Higgs mass depends on a time-dependent vacuum expectation value (VEV) of a scalar field, $\phi$\,.
This scalar evolves and eventually halts at a value rendering the effective Higgs mass much smaller than the cutoff.
This is achieved due to
 the fact that the potential of $\phi$ consists of a backreaction term that is switched on once the Higgs mass square gets negative and electroweak symmetry breaking (EWSB) is induced.
When compared with conventional models of naturalness, this class of models leads to a completely different phenomenology as there is no analog of top or gauge partners that can be discovered at colliders. Instead, as we will discuss in detail in this paper, for  experimental verification of the relaxion mechanism over a broad mass range we need to go to the low energy, high precision frontier.

Let us first present a very brief review of the relaxion mechanism. In relaxion models the value of $\mu^2$, the  mass squared term in the Higgs potential, changes during the course of inflation as it varies with the classical value of $\phi$,
\bea
V(H,\phi)&=&\mu^2(\phi) H^\dagger H+\lambda (H^ \dagger H)^2  \, ,\label{mu1}\\
\mu^2(\phi)&=&-\Lambda^2+g \Lambda \phi + \ldots    \, ,
\label{mu2}
\eea
which slowly rolls because of a potential,
\begin{eqnarray}
V(\phi)=r g\Lambda^3 \phi+\ldots ~  \,.
\label{roll}
\end{eqnarray}
In these equations $g$ is a coupling\footnote{Note that the coupling $g$ defined here is dimensionless and is related to $g_{\rm GKR}$, the one in ref.~\cite{kaplan},  via $g_{\rm GKR}=g \Lambda$.}   and $\Lambda$ is the  scale where the Higgs quadratic divergence gets cut off. Note that the operator in \eq{roll} can be radiatively generated by closing the Higgs loop in the term $g \Lambda \phi (H^ \dagger H)$ in \eq{mu1} and thus technical naturalness demands $r\gtrsim{1}/{16 \pi^2}$. In canonical models the field $\phi$ slowly rolls down (during inflation) from some initial large field value $\phi >{ \Lambda }/{g}$, such that $\mu^2$ is positive and the electroweak symmetry unbroken. It stops rolling shortly after the point $\phi_c\simeq {\Lambda}/{g}$, where $\mu^2$ becomes negative, electroweak symmetry is broken and the Higgs gets a vacuum expectation value,  $v^2(\phi)={-\mu^2(\phi)}/{\lambda}$. A crucial ingredient of the relaxation proposal is the feedback mechanism that  triggers a backreaction potential once the Higgs gets a VEV,
  \begin{eqnarray}
   \Delta V_{\rm br}(h,\phi)=-{\tilde{M}}^{4-j} \hat{h}^j \cos \left( \frac{\phi}{f} \right),
   \label{br}
  \end{eqnarray}
  where $1\leq j \leq 4$ is an integer and $\hat{h}=(v+h)/\sqrt{2}$.\footnote{For an alternative proposal where the rolling is stopped due to particle production see ref.~\cite{Hook:2016mqo}.}
  As $\phi$ continues rolling, $|\mu^2(\phi)|$ becomes larger, resulting in a monotonically increasing Higgs VEV and thus  increasing the backreaction's amplitude. Eventually the barriers become large enough and the relaxion stops rolling  at an arbitrary ${\cal O}(1)$ value of the phase $\phi_0/f$,
  \begin{eqnarray}
 \frac{\partial V(h,\phi)}{\partial \phi}&=&rg \Lambda^3+\frac{\Lambda^4_{\rm br}(v(\phi_0))}{f} \sin \left( \frac{\phi_0}{f} \right)=0 \ \Rightarrow \ \Lambda = \left|\frac{\Lambda^4_{\rm br}(v(\phi_0))}{rgf}  \sin \left( \frac{\phi_0}{f} \right)\right|^{1\over3}\,,\qquad
 \label{deriv}
  \end{eqnarray}
 where
 \begin{equation}\label{defLambr}
 \Lambda_{\rm br}^4(v(\phi_0))\equiv\tilde{M}^{4-j} v(\phi_0)^j/\sqrt{2}^j\,.\end{equation}
 Note that for values of $\Lambda^4_{\rm br}(v(\phi_0))\gg r g \Lambda^3 f$, \eq{deriv} is  satisfied for  $|\sin (\phi_0/f)|\ll 1$, but as ${\tilde{M}}^{4-j} v^j$ grows monotonically and the rolling starts at a random phase, $\phi/f$, the relaxion stops well before it reaches this stage. Therefore, generically the  phase $\phi_0/f$ has an  ${\cal O}(1)$ value. It is basically the result of a balance between the two terms controlling the derivative of the potential in \eq{deriv}. We shall return to this point when discussing CP-violation.  For a small enough value of $g$, the cut-off can be raised much above the electroweak scale. Such a small value of  $g$
 can be radiatively stable as in the $g\to 0$ limit we recover the discrete symmetry,
\begin{eqnarray}
\phi\rightarrow \phi+2\pi k f \,,\qquad k\in \mathbb{Z} \,. \label{discrete}
\end{eqnarray}
Higher dimensional terms such as $g^2 \phi^2 \Lambda^2, g^3 \phi^3 \Lambda\ldots $ contribute at the same order for $\phi\simeq \phi_c \simeq \frac{\Lambda}{g}$ and thus do not affect the above analysis. For the same reason, in  variants of the above model where only even powers of $\phi$ appear one can proceed along the same lines to obtain essentially the same results.

\enlargethispage{\baselineskip}

It is important to emphasize that in order to achieve higher values of the cut-off $\Lambda$, higher values of  $\Lambda_{\rm br}$ are required. Let us review the three main reasons for this. First of all, cosmological considerations during inflation (classical rolling must dominate over  quantum fluctuations, see ref.~\cite{kaplan})  put an upper bound on the cut-off which decreases if $\lbr$ is smaller,
\begin{eqnarray}
\Lambda \lesssim \Lambda_{cq}= \left(\frac{\Lambda_{\rm br}^{4} }{f}\right)^{1\over 6} \sqrt{M_{\rm Pl}}\,,
\label{cosmo}
\end{eqnarray}
where from now onward, for simplicity, by $v$, $\tilde{M}$ and $\Lambda_{\rm br}\,$ we will refer to the final values of these quantities at the relaxion minima $\phi=\phi_0$.
 An even stronger bound can be derived if we demand that the relaxion does not have transplanckian excursions,
 \begin{eqnarray}
 \Delta \phi\sim \frac{\Lambda}{g}\leq M_{\rm Pl}\Rightarrow \Lambda \lesssim \Lambda_{tp}= \left(\frac{M_{\rm Pl}}{ rf }  \right)^{1\over 4}\Lambda_{\rm br} \, ,
 \label{trp}
 \end{eqnarray}
 which once again favours a large $\lbr$.  As the requirement of subplanckian field excursions depends on quantum gravity assumptions and can be possibly evaded by UV model building, we will not take this as a strict bound and extend our analysis also to the transplanckian region. Finally, as argued in ref.~\cite{familon}, if the relaxion is a compact field, as any pseudo Nambu-Goldstone boson (PNGB) must be,    the ratio of the distance the relaxion rolls to the periodicity of the backreaction,
\begin{eqnarray}
n\simeq\frac{\Lambda/g}{f}\label{ndef}
\end{eqnarray}
 is, for a single axion sector, generically expected to be an  ${\cal O}(1)$ number. On the other hand, this ratio must be large to raise the cut-off substantially above the weak scale,  as \eq{deriv} implies,
\begin{eqnarray}
\Lambda\simeq\left(\frac{n}{r}\right)^{1/4}\Lambda_{\rm br}\,.
\label{en}
\end{eqnarray}
Thus smaller values of $\lbr$ require larger values of $n$ for a given cut-off scale.

\newpage

In this work we derive several new results relevant  for relaxion phenomenology.
We emphasise the importance of relaxion-Higgs mixing that is expected in a large class of relaxion models and focus on its experimental and observational implications.
As the relaxion-Higgs mixing turns out to be proportional to $\lbr^4$, these observational constraints put an upper bound on the backreaction scale, $\lbr$.
We also derive a theoretical upper bound on the backreaction scale $\Lambda_{\rm br}$. We further consider multiaxion (clockwork) models
 where  $n\sim e^N$, $N$ being the number of sites in these models, and show that for a too large value of $N$, the clockwork construction becomes tuned unless further structure is assumed. By eqs.~(\ref{cosmo})--(\ref{en}) we see that together these considerations favour lower values of the cut-off scale.

 In the following section we derive the expressions for the relaxion-Higgs mixing and in section~\ref{secbr} review existing backreaction models and bounds on the backreaction scale.
 In section~\ref{sectheory} we consider bounds on models with compact relaxions.  We find a rich variety of experimental and observational probes for the relaxion in the mass range 0.1 $\mu$eV to 50\,GeV described in detail in section~\ref{lab} and section~\ref{cosmosec}.  All our bounds are equally applicable to general Higgs portal models. As the relaxion  couplings to SM particles via the mixing are like that of a CP-even scalar,  in the sub-eV range  fifth force experiments can constrain large parts of the relaxion parameter space.  In the keV-MeV range  constraints on the relaxion parameter space arise from astrophysical star cooling bounds and cosmological probes of late decays, including constraints from entropy injection, BBN  observables, CMB distortions and distortion of the extragalactic background light (EBL) spectrum. In the MeV-GeV region we find that the most important bounds arise from cosmological entropy injection and BBN  bounds, cooling rate of the SN 1987  supernova, beam dump experiments and from constraints on rare $B$- and $K$-meson decays. Finally for GeV scale masses the  bounds arise from LEP Higgs-strahlung data and LHC Higgs coupling bounds on the $h \to \phi \phi$ channel. We also discuss how presently unconstrained parts of the relaxion parameter space would be probed by future data from experiments such as  the  PIXIE detector for CMB distortions, the NA62 experiment and especially the  SHiP beam dump experiment. In section~\ref{sect:summaryplots}  we  discuss the implications of our bounds on the theoretical parameter space of relaxion models. In section~\ref{cpv} we briefly discuss how the characteristic CP violation of relaxion models can be probed and finally  we conclude in section~\ref{conc}. Useful relations are derived in the appendices.

\enlargethispage{\baselineskip}

\section{Relaxion-Higgs mixing}
\label{secmix}
Relaxion models contain two sources of breaking of the shift symmetry, the one that allows the Higgs mass to scan as the relaxion rolls and the backreaction term.
In this section we will see how   the presence of both terms can lead to spontaneous CP-violation in the backreaction sector and a measurable relaxion-Higgs mixing (see also ref.~\cite{Kobayashi:2016bue}).
The full relaxion  potential is given by combining the terms in \eq{mu2}, \eq{roll} and \eq{br},
\begin{eqnarray} \label{heavy}
V= \left[- \Lambda^2 +g  \Lambda \phi+\dots\right] \hat{h}^2 -\tilde{M}^{4-j} \hat{h}^j \cos \left( \frac{\phi}{f} \right)+\lambda \hat{h}^4
  + rg \Lambda^3  \phi+ \dots \label{eq:V}\,.
  \end{eqnarray}
To obtain the mixing terms we expand around the minima of the fields $\phi$ and $H$,
 \begin{eqnarray}
\phi=\phi_0+ {\phi'}~~~~~~H^T=\left(0\,,~~\frac{v_H+h'}{\sqrt{2}}\right)\,.
\label{gen}
\end{eqnarray}
In models with even $j$, $v_H=v=246$\,GeV\@.  On the other hand,  as we will see in section~\ref{secbr}, the backreaction sector breaks electroweak symmetry in models with odd $j$. In these models, therefore,  $v_H=\sqrt{v^2-v'^2}$, where $v'$ is the electroweak symmetry breaking (EWSB) VEV in the backreaction sector.  The minimisation conditions and explicit mass matrices, $M^2_{ij}$, for  the $j=1$ and 2 cases can be, respectively,  found in appendix~\ref{app1} and~\ref{app2}. We find in both cases that the leading contribution to the mass matrix elements can be written entirely  in terms of the parameters of the backreaction sector. In particular,
\begin{equation}
{M^2_{h'\phi'} \over M^2_{h'h'}}={\cal O}\left({\lbr^4 \over m_h^2v_Hf}\right) \,, \  {M^2_{\phi'\phi'} \over M^2_{h'\phi'}}={\cal O}\left({v_H\over f}\right)\,.
\end{equation}
for both $j=1$ and $j=2$. In addition, as discussed below, we expect $\lbr\lesssim v_H$
which implies that the relaxion-Higgs mixing angles is naturally small, $\sin \theta \ll 1$.
 We find that, to leading order the relaxion Higgs-mixing angle $\theta$ and the relaxion mass $\mphi$ are,
\begin{eqnarray}
\sin \theta &\simeq& \tan \theta\simeq\frac{M^2_{h'{\phi'}}}{M^2_{h'h'}}= j\frac{\lbr^4}{v_H f  m_h^2} \sin \left(\frac{\phi_0}{f}\right)\label{eq:sintj}\,,\\
m_\phi^2 &\simeq& \frac{\lbr^4}{ f^2}\left(\cos \left(\frac{\phi_0}{f}\right)-\frac{j^2\lbr^4 }{v_H^2 m_h^2}\sin^2 \left(\frac{\phi_0}{f}\right)\right).
\label{massmixj}
\end{eqnarray}
As anticipated, the mixing angle is proportional to the spontaneous CP-violating spurion $\sin (\phi_0/f)$ in the backreacting sector. In more complicated relaxion models  there can be mechanisms to suppress this phase. An example is the model with the QCD axion where a small phase is necessary to be compatible with the non-observation of a strong CP-phase. In such models (see also ref.~\cite{bcn} and ref.~\cite{DiChiara:2015euo}), relaxion-Higgs mixing is also suppressed.

\paragraph{Couplings:} As the relaxion mixes with the Higgs boson, it inherits its couplings to SM particles suppressed by a the mixing angle $\sin \theta$ as a universal factor --- such as in Higgs portal models.\footnote{Note that in $j=1$ models the Higgs couplings themselves might differ from their SM values because of the reduced Higgs VEV, $v_H=\sqrt{v^2-v'^2}$; in the following we will assume that $v'\lesssim 100$\,GeV so that these are at most 10 $\%$ effects which we would ignore (see also section~\ref{secbr}).} For $g_{\phi \psi}$, the coupling to pairs of matter fields $\psi$, and $g_{\phi V}$, the coupling to pairs of $V=W^\pm$ or $Z$, the couplings are  given by
\begin{eqnarray}
g_{\phi f, \phi V } = \sin \theta g_{h f, h V} \,. \label{eq:gphipsi}
\end{eqnarray}
At the loop level, the relaxion couples via quark loops to gluons and quark  and $W^\pm$ loops  to photons,
\begin{eqnarray}
{\cal L} \supset
           - \frac{g_{\phi \gamma}}{4}  \phi F^{\mu\nu}F_{\mu\nu}
           - \frac{g_{\phi g} }{4}  \phi G^{\mu\nu}G_{\mu\nu}\,,\\
\end{eqnarray}
where
\begin{eqnarray}
g_{\phi g}&=&\frac{ \alpha_{s}\sin \theta }{4 \pi v}\left| \sum_{\rm fermions} N_{c,f} Q_f^2 A_F(\tau_f)\right|\nonumber\,,\\
g_{\phi \gamma}&=&\frac{ \alpha_{em}\sin \theta }{2\pi v}\left| A_W(\tau_W)+\sum_{\rm fermions} N_{c,f} Q_f^2 A_F(\tau_f)\right|\, \label{gphigam}
\end{eqnarray}
with
\begin{eqnarray}
A_F(\tau)&=&\frac{2}{\tau^2}(\tau+(\tau-1)f(\tau))\,\\
A_W(\tau)&=&-\frac{1}{\tau^2}\left(2 \tau^2+3 \tau+ 3(2\tau-1)f(\tau)\right)\,\\[2mm]
f(\tau) &=& \left\{
\begin{array}{ll}
\arcsin^2\sqrt{\tau} & \tau\leq 1\,\\[3mm]
\displaystyle -\frac{1}{4}\left[\log\frac{1+\sqrt{1-\tau^{-1}}}{1-\sqrt{1-\tau^{-1}}}- i \pi\right]^2 & \tau > 1\,
\end{array}
\right.
\end{eqnarray}
where $\tau_x = m^2_h / 4 m_x^2$.

Let us finally comment on the  pseudoscalar couplings of the relaxion to Standard Model particles, as these may have a significant impact on the experimental probes discussed in the following.
However, these couplings are model-dependent and as the relaxion potential can be controlled by a sequestered sector~\cite{kaplan,familon} these couplings could be in principle suppressed relative to the ``Higgs-portal" couplings discussed above (which are at the core of the relaxion construction). As we show in appendix~\ref{ps}, this is  the case in existing backreaction models (see section~\ref{secbr}) where we find that these couplings are in magnitude  generally smaller than or equal to the Higgs portal couplings.   An exception is the  pseudoscalar coupling to photons which in some backreaction models (see appendix~\ref{ps}) can be larger than the one induced via Higgs mixing while in other models is of  the same size as the Higgs-portal coupling.  As the presence of  a large pseudoscalar coupling to photons is thus  model-dependent, we will comment on its implications only   qualitatively.

\section{Review of backreaction models and existing bounds on \texorpdfstring{\boldmath $\Lambda_{\rm br}$}{Lambda(br)}}
\label{secbr}
As both the cutoff of the theory as well as the relaxion-Higgs mixing depends polynomially on the back reaction scale, it is important to examine what is its allowed range.
 In this section we thus describe  the different backreaction models in the literature  and discuss  various bounds on the size of the scale $\tilde{M}$ or $\Lambda_{\rm br}$ that appears in the backreaction potential [see \eq{defLambr}].

  Note, first of all, that  for odd $j=1$ or 3,  a non-zero $\tilde{M}$ in \eq{br} must break electroweak symmetry which already suggests $\tilde{M}\lesssim v$, but let us analyze this case in more detail. The simplest relaxion model~\cite{kaplan} where the backreacting sector is QCD and the relaxion  couples to gluons like the axion, $ \frac{\phi}{f} G_{\mu \nu} \tilde{G}^{\mu\nu}$
 is an example of a $j=1$ model.  Non-perturbative effects generate a potential for the axion,
  \begin{eqnarray}
 \Delta V_{\rm br}\simeq -m_u \cos \frac{\phi}{f}  \langle\bar{q} q\rangle \simeq -4 \pi f_\pi^3 m_u\cos \frac{\phi}{f},
  \end{eqnarray}
  where $\Lambda_{\rm br}^4= 4 \pi f_\pi^3 m_u = 4 \pi f_\pi^3 y_u v/\sqrt{2}$  is set by the pion decay constant $f_\pi$ and the up quark mass. As the relaxion stops at a generic value of the phase, $\phi/f$, QCD relaxion models are generally in conflict with the non-observation of a large value of the strong CP phase.  This problem can be solved in more complicated variants where there is a dynamical mechanism to make the above phase small.

  An alternative approach would be to give-up the solution to the strong CP problem and to consider an additional strong sector. For instance,  a new technicolor-like strong sector would lead to an EWSB condensate of techniquarks,
$\langle \bar{U}_L U_R+\bar{D}_L D_R\rangle \simeq v'^3$, where $U_{L,R}$ and $D_{L,R}$ are quarks with the same electroweak charges as  the SM quarks $u_{L,R}$ and $d_{L,R}$, but charged under the new strong group and not QCD\@. If the relaxion is coupled to the operator  $G'_{\mu \nu} \tilde{G'}^{\mu\nu}$, involving the strong sector gauge bosons ($G'_{\mu \nu}$ corresponds to the new strong sector field strength), a backreaction  is generated with $j=1$ and
\begin{eqnarray}
 \Lambda_{\rm br}^4\simeq \frac{y v'^3v_H}{\sqrt{2}}\,,
 \end{eqnarray}
where $y$ is the  smaller of the $U$  or $D$ Yukawa coupling with the SM Higgs, and $v_H$ is the VEV of the Higgs doublet so that
$v'^2+v_H^2=v^2=(246 {\rm~GeV})^2$. Such a scenario is constrained by Higgs and electroweak (EW) precision observables as Higgs couplings deviate from SM values by ${\cal O}(v'^2/v_H^2)$. Requiring these deviations to be smaller than 20$\%$ gives, $v'\lesssim 100 {\rm~GeV}$. Together with this upper bound and the fact that the quarks must not have too large an explicit mass, i.e.\ we must have $y v_H \ll 4 \pi v'$ and hence $y \lesssim 1$, we obtain an upper bound on $\Lambda_{\rm br}$,
 \begin{eqnarray}
\Lambda_{\rm br}\lesssim 100 {\rm~GeV}.
 \end{eqnarray}
 It is worth pointing out that such models would not be as strongly constrained as typical technicolor models because the condensate does not need to explain the large top mass and because the presence of an elementary Higgs somewhat alleviates the tension with electroweak precision observables~\cite{yless}. In this work we have assumed $v'\lesssim 100$\,GeV and ignored  ${\cal O}(v'^2/v_H^2)$ effects.

A  less constrained model with $j=2$  was presented in ref.~\cite{kaplan}. In this model, $\phi$ couples to  $G'_{\mu \nu} \tilde{G'}^{\mu\nu}$,  the gauge bosons of an EW symmetry preserving strong sector. The Higgs couples to two vector-like leptons charged under this strong group as follows,
  \begin{eqnarray}
  {\cal L}=y_1 LHN+y_2 L^cH^\dagger N^c -m_L L L^c-m_N N N^c+{\rm h.c.}\,,
  \label{lagax}
  \end{eqnarray}
 where $(L, N)$ have the same quantum numbers as the SM lepton doublet and   right-handed neutrino, respectively, and $(L^c, N^c)$ are in the conjugate representations. If we take $m_N\ll 4 \pi f_\pi' \ll m_L$, only the fermion $N$ forms a  condensate that is EW preserving.  Upon integrating out $L, L^c$,  the Higgs contributes to the mass of $N$ as follows,
$ \Delta m_N= y_1 y_2 \hat{h}^2 /m_L$ so that  the relaxion potential gets the backreaction
\begin{eqnarray}
 \Delta V_{\rm br} \simeq - 4 \pi f_{\pi'}^3 \Delta m_N \cos \frac{\phi}{f}=-\frac{4 \pi f_{\pi'}^3 y_1 y_2\hat{h}^2 }{m_L}\cos \frac{\phi}{f}\,.
\label{brstr}
\end{eqnarray}

 A perturbative $j=2$ model was presented in ref.~\cite{familon}. In this model the relaxion is a familon, the PNGB of a spontaneously broken flavour symmetry. Let us consider the Lagrangian,
 \begin{eqnarray} \label{Lfamilon}
{\cal L}=y_1 e^{i\phi/f}LHN+y_2 L^cH^\dagger N -m_L L L^c-\frac{m_N}{2} N N+  {\rm h.c.}
\label{lagfam}
\end{eqnarray}
where $L$ and $L^c$ have the same quantum numbers as before, and $N$ is a SM singlet fermion. The one-loop Coleman-Weinberg potential of the relaxion $\phi$ reads
\begin{eqnarray}
\label{VCWbefore}
 \Delta V_{\rm br} \simeq  -\frac{1}{2 \pi^2} m_L m_N  y_1 y_2 \hat{h}^2  \cos \left(\frac{\phi}{f}\right)  \log\left( \frac{\Lambda^2}{\tilde{m}^2}\right) \,,
\end{eqnarray}
where $\tilde{m}$ is the larger of $m_L$, $m_N$.

 A theoretical challenge that any $j=2$ model faces is that at the quantum level the backreaction term [see Eq.~\eqref{br}] generates the term
 $\frac{\tilde{M}^2\Lambda_c^2}{16 \pi^2} \cos \frac{\phi}{f}$
upon closing the $\hat{h}$ loop. This term is independent of the Higgs VEV, which implies the presence of an oscillatory potential even before the Higgs condenses~\cite{bcn}. Thus, the relaxion stops rolling prematurely, before EWSB, unless the scale $\Lambda_c$ at which the Higgs loop is cut-off satisfies
 \begin{eqnarray}
 \Lambda_c\lesssim 4 \pi v\,.
 \label{2loop}
 \end{eqnarray}
  In axion-like models this is automatically satisfied because the instanton contribution are highly suppressed at energy scales larger than the confinement scale, $4 \pi f_{\pi'}$, so that  \eq{2loop} implies $f_{\pi'}\lesssim v$. In the model of Eq.~(\ref{lagax}) there is actually  another contribution to the potential that exists even before EWSB,  $\Delta V_N \simeq 4 \pi f_{\pi'}^3  m_N \cos \frac{\phi}{f}$, where technical naturalness requires that $m_N$ must be larger than $y_1 y_2  m_L \log(\Lambda/m_L)/(16 \pi^2)\,$. Demanding the above EW preserving contribution to be  smaller than the backreaction generated upon EWSB, $4 \pi f_{\pi'}^3 \Delta m_N \cos \frac{\phi}{f}$, we obtain $m_L\lesssim {4 \pi v}/{\sqrt{\log{\Lambda/m_L}}}$. Together with these bounds, the requirement $\Delta m_N \ll 4 \pi f_{\pi'}\ll m_L$ so that $N$ forms a condensate and $L$ does not, implies that in this model
\begin{eqnarray}
\frac{1}{16 \pi^2}\left(\frac{y_1 y_2 v}{8 \pi}\right)^4\ll\Lambda^4_{\rm br} \ll 16 \pi^2 v^4\,,
\end{eqnarray}
where we have assumed $m_L\gtrsim v$ due to experimental bounds for the first inequality. In the perturbative familon case, of Eq.~(\ref{lagfam}), a simple extension can ensure that the constraint in \eq{2loop} is satisfied as followed. The Majorana mass $m_N$ is actually induced via a mini see-saw mechanism. A new heavier fermion $N^c$ is added to the theory,
 \begin{eqnarray}
 {\cal L}\supset-m_D N N^c-\frac{m_{N^c}}{2} N^c N^c\,.
 \end{eqnarray}
 After $ N^c$ is integrated out, the Majorana mass of $N$ is induced, $m_N=m_D^2/m_{N^c}$. One can show in this case that two-loop corrections to the relaxion potential do not get contributions from energies above the scale $m_{N^c}$ so that we get $\Lambda_c=m_{N^c}$, and \eq{2loop} is satisfied as long as $m_{N^c}\lesssim4 \pi v$. As $m_{L,N} \ll m_{N^c}$, this implies $m_{L,N} \ll 4 \pi v$, and thus
\begin{eqnarray}
\Lambda^4_{\rm br} \ll 64  \pi^2 v^4
\end{eqnarray}
where we have assumed  $y_{1,2}< 4 \pi$.

Now let us discuss some model-independent bounds on the backreaction scale. First note that \eq{massmixj} implies that for a non-tachyonic $\phi$,
\begin{eqnarray}
\lbr^2<(\lbrmax)^2=\frac{m_h v}{j}\frac{\sqrt{\cos \left({\phi_0}/{f}\right)}}{\sin \left(\phi_0/f\right)}\,.
\label{tach}
\end{eqnarray}
 Finally notice that in the presence of the mixing the Higgs-like eigenvalue would satisfy,
$m_h^2 >M^2_{h'h'}\,$. For the $j=1$ case this leads to a  bound on $\tilde{M}$ simply arising from  the expression for the Higgs mass that is given by (see Eq.~\eqref{mincon})
\begin{eqnarray}
m_h^2\geq\frac{\tilde{M}^3  \cos(\phi_0/f)}{\sqrt{2}v_H}+2\lambda v_H^2\,
 \label{mhone}
\end{eqnarray}
where the inequality becomes an equality in the limit of no relaxion-Higgs  mixing. We must have $\lambda>0$ to ensure that the potential does not have a runaway direction which implies the following bound,
 \begin{eqnarray}
 \tilde{M}^3\lesssim  \frac{\sqrt{2} m_h^2 v_H}{\cos(\phi_0/f)} \Rightarrow \Lambda_{\rm br}^4\lesssim \frac{m_h^2 v_H^2}{\cos(\phi_0/f)}\,.
 \end{eqnarray}

\section{New bounds on compact relaxions}
\label{sectheory}

 In this section we consider simple multiaxion (clockwork) models   and then show that these suffer from stability issues when the number of sites (axions), $N$, becomes too large. The instability is, in fact, related to the very same issue of highly irrelevant operators that plagues the two-site construction in ref.~\cite{familon}. First of all note that in realistic relaxion models the coupling $g$ in \eq{mu2} and  \eq{roll} is obtained from a compact term (at least in QFT constructions where the relaxion is a pNGB), but with a larger periodicity $F$~\cite{familon},
\begin{eqnarray}
V(\hat{h},\phi)=\left(\kappa \Lambda^2-\Lambda^2 \cos \left(\frac{\phi}{F}+\alpha\right)\right)\hat{h}^2-r \Lambda^4\cos \left(\frac{\phi}{F}+\alpha\right)-\Lambda_{\rm br}^4 \cos \frac{\phi}{f}\, ,
\label{relpot}
\end{eqnarray}
 which allows us to make the following identifications for $\kappa<1$:
\beq
g={\Lambda\over F}\,, \ \ n={F\over f}\,.\label{compact}
\eeq
One can now directly obtain  \eq{en} by demanding $V'(\phi)=0$ using \eq{relpot},
\beq
\Lambda\simeq \left(\frac{n}{r}\right)^{1/4} \Lambda_{\rm br}.
\label{en2}
\eeq

 As shown in ref.~\cite{choiim} and~\cite{kaplan2}, the Choi-Kim-Yun (CKY) alignment mechanism~\cite{cky}\footnote{The mechanism was proposed as a generalization of the Kim-Nilles-Peloso alignment mechanism~\cite{Kim:2004rp} from 2-axion to an $N$-axion-alignment.} (also known as the clockwork mechanism in the relaxion context) for multiple axions (or PNGBs) can provide a relaxion potential having two periodicities with a large ratio $F/f\sim e^N$, $N$ being the number of axions.  Let us first review these multiaxion models.  We describe here the realization of ref.~\cite{kaplan2}.  Consider $N+1$ complex scalar fields $\Phi_1$ to $\Phi_{N+1}$ with the~potential
\begin{eqnarray}
V(\Phi)=\sum^N_{j=1}\left(-m^2 \Phi_i^\dagger \Phi_i+\frac{\lambda}{4}|\Phi_i^\dagger \Phi_i|^2\right)+\epsilon\left(\Phi_1^\dagger \Phi_2^3+\Phi_2^\dagger \Phi_3^3\ldots .\Phi_{N}^\dagger \Phi_{N+1}^3+{\rm h.c.}\right)
\label{clockwork}\,.
\end{eqnarray}
The above potential respects a  $\U(1)$ symmetry under which the fields $\Phi_1,\Phi_2\ldots \Phi_N$ have charges $Q=1,\frac{1}{3}\ldots ,\frac{1}{3^N}$. For simplicity the symmetry preserving cross terms such as $\Phi_i^\dagger \Phi_i \Phi_j^\dagger \Phi_j$ have been ignored and an approximate permutation symmetry has  been assumed (for $\epsilon\to 0$) so that the masses $m^2$ and quartic couplings $\lambda$ are  equal for all the fields. For $\epsilon \ll \lambda$  the  radial parts of the fields obtain a VEV,  $\Phi_i=\frac{\hat{f}}{\sqrt{2}} e^{i \phi_i/\hat{f}}$ where  $\hat{f}^2=4 m^2/\lambda$, such that at low energies only the angular degree of freedom remains. $N$ superpositions of the angular fields obtain masses, but the direction
\begin{eqnarray}
\phi_0=\frac{1}{\cal N}\left(\phi_1+\frac{\phi_2}{3}+\frac{\phi_3}{9}\ldots +\frac{\phi_{N+1}}{3^{N}}\right)
\label{vec}
\end{eqnarray}
is a flat direction that describes a Goldstone boson. The Goldstone mode  has an ${\cal O}(1)$ overlap with the first site and is exponentially suppressed overlap with the last site,
\begin{eqnarray}
\langle\phi_1 |\phi\rangle &=&\frac{1}{\cal N}\,,\label{eq:overlap1}\\
\langle\phi_N |\phi\rangle &=&\frac{1}{3^{N}{\cal N}}\,,
\label{overlap}
\end{eqnarray}
where ${\cal N}=\sqrt{\sum_{j=1}^{N+1} \frac{1}{3^{2 (j-1)}}}$ is the norm of the vector defined by \eq{vec}. Let us now introduce some anomalous breaking of the global $\U(1)$ at the first and last sites,
\begin{eqnarray}
&&\left(\frac{\phi_{N+1}}{\hat{f}} +\theta \right) G_{\mu \nu} \tilde{G}^{\mu\nu}+\left(\frac{ \phi_1}{\hat{f}} +\theta'  \right)G'_{\mu \nu} \tilde{G}'^{\mu\nu}\,,\nonumber\\
&\to\,&
\left(\frac{\phi_{0}}{f} +\theta \right) G_{\mu \nu} \tilde{G}^{\mu\nu}+\left(\frac{ \phi_0}{3^{N} f} +\theta'\right)G'_{\mu \nu} \tilde{G}'^{\mu\nu}
\label{expl}\,,
\end{eqnarray}
where $f={\cal N} \hat{f}$ and we have used eqs.~(\ref{eq:overlap1}),~(\ref{overlap}) to rewrite the first line in terms of $\phi_0$. Non-peturbative effects now generate the desired  relaxion potential in \eq{relpot} with $F=3^{N} f$ so that \eq{en2} now becomes
\begin{equation}
 n=3^{N}=\frac{r \Lambda^4}{\lbr^4}\,.\label{en3}
\end{equation}
Thus we see that the CKY/clockwork mechanism can give us a cut-off that grows exponentially with the number of axions, $N$.  Note that  the above analysis holds only if
 \begin{eqnarray}
  r\Lambda^4 \ll \frac{\epsilon f^4}{4}\,,
 \label{claudiabnd}
 \end{eqnarray}
 so that  the potential generated by the anomalous breaking of the $\U(1)$ symmetry in \eq{relpot} is subdominant compared to the potential generated from \eq{clockwork}, the linear combination in   \eq{vec} remains a Goldstone mode, and all the heavier modes can be~decoupled.

 We now show that,  for a too large $N$, these models become finely tuned if we relax the approximate permutation symmetry in \eq{clockwork} that was assumed only for convenience of calculation. If we relax this assumption some of the mass square terms might be positive. Let us assume, for instance, that $k-1$ consecutive fields, $ \Phi_{{n_1}+1}\ldots \Phi_{{n_1}+k-1}$
 have  positive mass square terms so that there are no corresponding PNGB modes  $\phi_{{n_1}+1}\ldots \phi_{{n_1}+k-1}$ for these scalars.
 At first sight, this breaks the link in the axion chain because --- instead of one Goldstone mode as in \eq{vec} --- there are now two decoupled Goldstones fields [in the absence of the subdominant boundary terms of \eq{expl}],
\begin{eqnarray}
\phi_{01}&=&\frac{1}{{\cal N}_1}\left(\phi_1+\frac{\phi_2}{3}+\ldots \frac{\phi_{n_1}}{3^{n_1-1}}\right)\,,
\label{vec1}\\
\phi_{02}&=&\frac{1}{{\cal N}_2}\left(\phi_{n_1+k}+\frac{\phi_{n_1+k+1}}{3}+\ldots \frac{\phi_{N}}{3^{n_2}}\right)\,,
\label{vec2}
\end{eqnarray}
where $n_2=N-n_1-k$ and ${\cal N}_1$ and  ${\cal N}_2$ are again normalisation constants.   None of the above modes can be identified with the relaxion as no single  mode above is subject to both the backreaction at the first site and the rolling potential at the last site.  However, the link between the two chains is not completely lost, as a process like $\Phi_{n_1-1} \to 3 \Phi_{n_1}\to\ldots  \to 3^m \Phi_{n_1+k}$
generates a higher dimensional operator that weakly couples the two sectors,
\begin{eqnarray}
\epsilon^{\frac{3^k-1}{2}} \frac{\Phi_{n_1}^\dagger \Phi^{3^k}_{n_1+k} }{m^{3^k-3}} &\to& \varepsilon \frac{\epsilon \hat{f}^4}{2} \cos \left( \frac{3^k\phi_{n_1+k}}{\hat{f}}-\frac{\phi_{n_1}}{\hat{f}}\right)\,, ~~~~{\rm where~~~}\varepsilon=\left(\frac{\epsilon}{\lambda}\right)^{\frac{3^k-3}{2}}
\end{eqnarray}
is an exponentially small number due to $\epsilon/\lambda \ll1$.  More precisely, as the $N-1$ heavier pseudo-Goldstone modes that have masses $m^2_\phi \sim 3^2 \epsilon \hat{f}^2/2$ (see ref.~\cite{kaplan2}), must be much lighter than the radial modes $\rho$ with $m^2_\rho \sim \lambda \hat{f}^2/2$, one needs
\begin{eqnarray}
\frac{m^2_\phi }{m^2_\rho}\lesssim \frac{1}{9} \Rightarrow  \frac{\epsilon }{\lambda}\lesssim \frac{1}{3^4}  \Rightarrow \varepsilon \lesssim 3^{-2(3^k-3)}\,.
\end{eqnarray}
 With the above term in the potential we once again obtain that $\phi_0$ from \eq{vec} is a Goldstone mode. With the addition of the explicit breaking terms on the first and last site in \eq{expl}, the terms relevant for the potential of $\phi_{01}$ and $\phi_{02}$ are,
\begin{eqnarray}
V(\phi_i)&\supset&-\Lambda_{\rm br}^4 \cos\left(\frac{\phi_1}{\hat{f}}+\theta\right)+\varepsilon \frac{\epsilon \hat{f}^4}{2} \cos \left(\frac{3^k\phi_{n_1+k}}{\hat{f}}-\frac{\phi_{n_1}}{\hat{f}}\right)- r\Lambda^4\cos\left(\frac{\phi_N}{\hat{f}}+\theta'\right)\nonumber\\
&=&-\Lambda_{\rm br}^4 \cos\frac{\phi_{01}}{{\cal N}_1\hat{f}}+\varepsilon  \frac{\epsilon \hat{f}^4}{2} \cos \left(\frac{3^k\phi_{02}}{{\cal N}_2\hat{f}}-\frac{\phi_{01}}{3^{n_1-1}{\cal N}_1\hat{f}}\right)- r\Lambda^4\cos\left(\frac{\phi_{02}}{3^{n_2}{\cal N}_2\hat{f}}+\alpha'\right)\nonumber\,,\\
\label{vare}
\end{eqnarray}
 where we have appropriately redefined $\phi_{01}$ and $\phi_{02}$ so that the phase appears only in the last term. The two lightest modes now are superpositions of \eq{vec1} and \eq{vec2}.  The mass matrix of $\phi_{01}$ and $\phi_{02}$ is given by
\begin{eqnarray}
M&=& \frac{\epsilon f^2}{2}\left(
\begin{array}{c@{\quad}c}
-\frac{\Lambda_{\rm br}^4}{{\cal N}^2_1\epsilon \hat{f}^4}+3^{-2(n_1-1)}\frac{\varepsilon}{{\cal N}_1^2} & 3^{k-n_1+1}\frac{\varepsilon}{{\cal N}_1{\cal N}_2} \\[3mm]
3^{k-n_1+1}\frac{\varepsilon}{{\cal N}_1{\cal N}_2} &   3^{2k}\frac{\varepsilon}{{\cal N}_2^2}- 3^{-2 n_2}\frac{r\Lambda^4}{ {{\cal N}_2^2}\epsilon \hat{f}^4}
\end{array}
\right) \, ,
\label{y3d}
\end{eqnarray}
which results in, up to normalization factors, the two mass eigenstates,
\begin{eqnarray}
\phi_{m1}&= c_\alpha\phi_{01} + s_\alpha \phi_{02}\,, \nonumber\\
\phi_{m2}&=-s_\alpha\phi_{01} + c_\alpha \phi_{02}\,,
\label{evecs}
\end{eqnarray}
where $s_\alpha= \sin \alpha$, $c_\alpha= \cos \alpha$ and $\alpha$ is the mixing angle. Let us first show that in the limit that contribution of the term proportional to $\Lambda^4$  to the gradient of the $\phi_{m2}$ potential is subdominant, i.e.\ $\Lambda^4 \ll 3^{n_2+k} \varepsilon \epsilon \hat{f}^4$, we recover the usual relaxion potential. In this limit we obtain $\tan \alpha= 3^{-n_1-k} {\cal N}_2/{\cal N}_1$, and the first eigenstate in \eq{evecs} becomes identical to  the relaxion mode in \eq{vec}.
To obtain the Lagrangian for the lightest mode we first
use the condition to stabilise $\phi_{m2}$\,, which in this limit reads
\begin{eqnarray}
\frac{\partial V}{\partial \phi_{m2}}\simeq-\frac{3^{k}c_\alpha}{{\cal N}_2}  \frac{\varepsilon\epsilon \hat{f}^3}{2} \sin \left(\frac{3^{k}c_\alpha}{{\cal N}_2}\frac{\phi_{m2}}{\hat{f}}\right)=0\,.
\label{minim2abc}
\end{eqnarray}
Substituting the solution $\langle \phi_{m2}\rangle=0$ in \eq{vare} and using $\tan \alpha= 3^{-n_1-k} {\cal N}_2/{\cal N}_1$ yields the potential in \eq{relpot}.

 In the opposite limit, i.e.~$\Lambda^4 \gg 3^{n_2+k} \varepsilon \epsilon f^4$, the gradient of the $\phi_{m2}$ potential  is dominated by the term proportional to $\Lambda^4$,
\begin{eqnarray}
\frac{\partial V}{\partial \phi_{m2}}\simeq\frac{\Lambda^4 c_\alpha }{3^{n_2}{\cal N}_2 \hat{f}}\sin\left(\frac{s_\alpha \phi_{m1}+ c_\alpha\phi_{m2}}{3^{n_2}{\cal N}_2\hat{f}} +\alpha'\right)=0\,.
\end{eqnarray}
which drives $\phi_{m2}$ to the global minimum of the rolling potential giving the Higgs an  ${\cal O}(\Lambda^2)$ mass. Therefore for the relaxion mechanism to work one needs $\Lambda^4 \ll 3^{n_2+k} \varepsilon \epsilon \hat{f}^4$, \hbox{which~implies}
 \begin{eqnarray}
r \Lambda^4 \ll 3^{-z}\lambda \hat{f}^4\,,
 \label{newbound}
 \end{eqnarray}
 where
 \begin{eqnarray}
 z=2(3^k-3)-N+n_1+4\,.
 \end{eqnarray}
 We see that for $k=4, \, \lambda=1, \, r={1}/{16 \pi^2}$ and $N=28$, using $f<\mpl$ we get  $\Lambda \ll 2$\,TeV for any positive integer $n_1$, so that the relaxion mechanism cannot even address the little hierarchy problem in this case.

 How long must the relaxion chain be so that a sequence of  $k-1=3$ consecutive positive masses  becomes highly probable? To compute this probability we need to find the number, ${\cal N}_3(N)$,  of   sequences of $N$ `+' or `-'  signs with at least one chain of 3 consecutive positive signs  `+++	' . It can be shown that ${\cal N}_3(N)$ obeys the following recursion relation,
 \begin{eqnarray}
 {\cal N}_3(N+1)=2.{\cal N}_3(N)+\left[2^{N-3}-{\cal N}_3(N-3)\right].
 \label{rec}
 \end{eqnarray}
 Here the first term comes from the fact that if we already have at least one `+++' chain  in a sequence of  $N$ axions, by adding either a `+' or `-'  sign at the $N+1$ th position we obtain  an arrangement of size $N+1$ satisfying our criterion. This does not include arrangements of  $N$ axions with no chain of 3 consecutive positive `+' signs  but having a `-++' at the end such that we get a required arrangement if at the $N+1$ th position we add a `+'  sign; this is taken care of by the second term in \eq{rec}. Finally in  the last term  we subtract the double counting resulting in cases where the sequence captured by the second term already includes a `+++' in the remaining subchain.

 To obtain a successful relaxion model, however, we are interested in an $N$-site sequence with no `+++' chain,  ${\cal N}'_3(N)$, which is given by
 \begin{equation}
 {\cal N}'_3(N)=2^N- {\cal N}_3(N)\,.
 \end{equation}
  It turns out that ${\cal N}'_3(N)$ satisfies the following familiar relation,
  \begin{eqnarray}
 {\cal N}'_3(N+1)={\cal N}'_3(N)+{\cal N}'_3(N-1)+{\cal N}'_3(N-2)\,,
 \label{rec2}
 \end{eqnarray}
 which is nothing but the recurrence relation of the 3-step Fibonacci sequence.\footnote{The $n$-step Fibonacci sequence $\rm{fib}_n$ is a sequence where any number in the sequence is the sum of the previous $n$ numbers.} By inspection, ${\cal N}'_3(3)=7\,$, the 5th element of the 3-step Fibonacci sequence so that we must have ${\cal N}'_3(N)={\rm fib}_3(N+2)$.
 Our arguments can be easily generalized to find the number of arrangements with no chains of $k-1$ positive masses which turns out to be just the $(k-1)$-step  Fibonacci sequence. Hence the probability to randomly obtain a sequence with at least one chain of  $k-1$ positive masses is
 \begin{eqnarray}
  {\cal P}(k-1,N)=1-\frac{{\rm fib}_{k-1}(N+2)}{2^N}\,.
 \end{eqnarray}
 We find that for $N\geq 28$ the probability of having at least $k-1=3$ consecutive positive masses in a chain of $N$ axions is ${\cal P}(k-1,N) \simeq 90 \%$. Thus for for $N\geq 28$, from \eq{newbound}  generically we have $\Lambda \ll 3$\,TeV as already discussed above.

 For  $N\lesssim 28$ axions there is the possibility of raising the cut-off to a value of
 \begin{eqnarray}
 \Lambda\lesssim 3^{28/4} (16 \pi^2 \lbr^4)^{1/4}=1000{\rm~TeV} \sqrt{\frac{\lbr^2}{m_h v}}\,,
 \label{clcut}
 \end{eqnarray}
 where we have used \eq{en3}  and the numerical value above is for  $\lbr \simeq \sqrt{m_h v}$. As we will show in sections~\ref{lab} and~\ref{cosmosec}, experimental probes can constrain $\lbr$ to even smaller values as a function of $f$ (or alternatively the relaxion mass) and this in turn would imply an even lower cut-off in accordance with \eq{clcut}.

\section{Laboratory  probes of relaxion-Higgs mixing}
\label{lab}

\begin{figure}
\centerline{
  \subfigure[]{\includegraphics[width=0.48\textwidth]{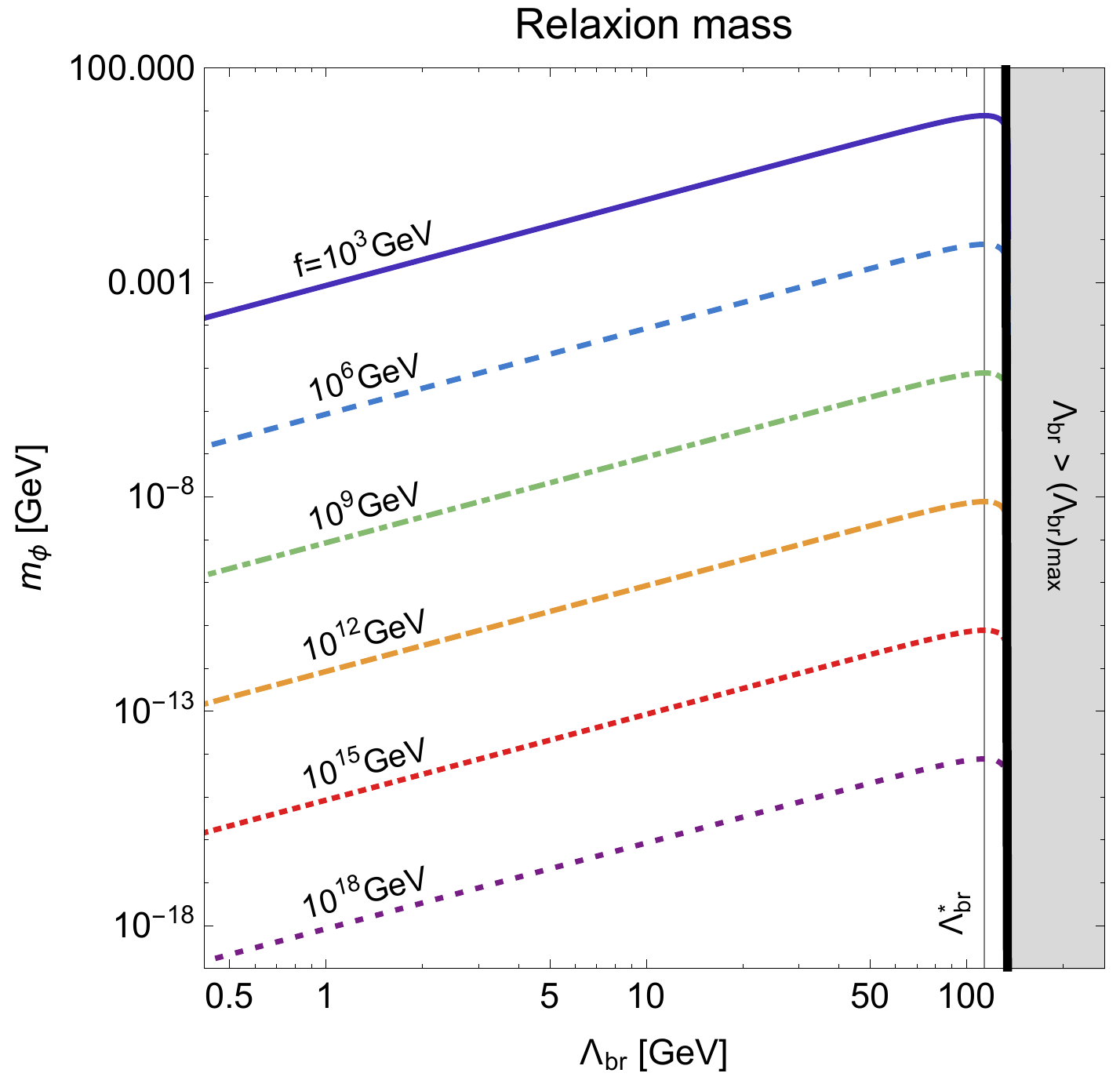} \label{fig:mass}}
  \subfigure[]{\includegraphics[width=0.46\textwidth]{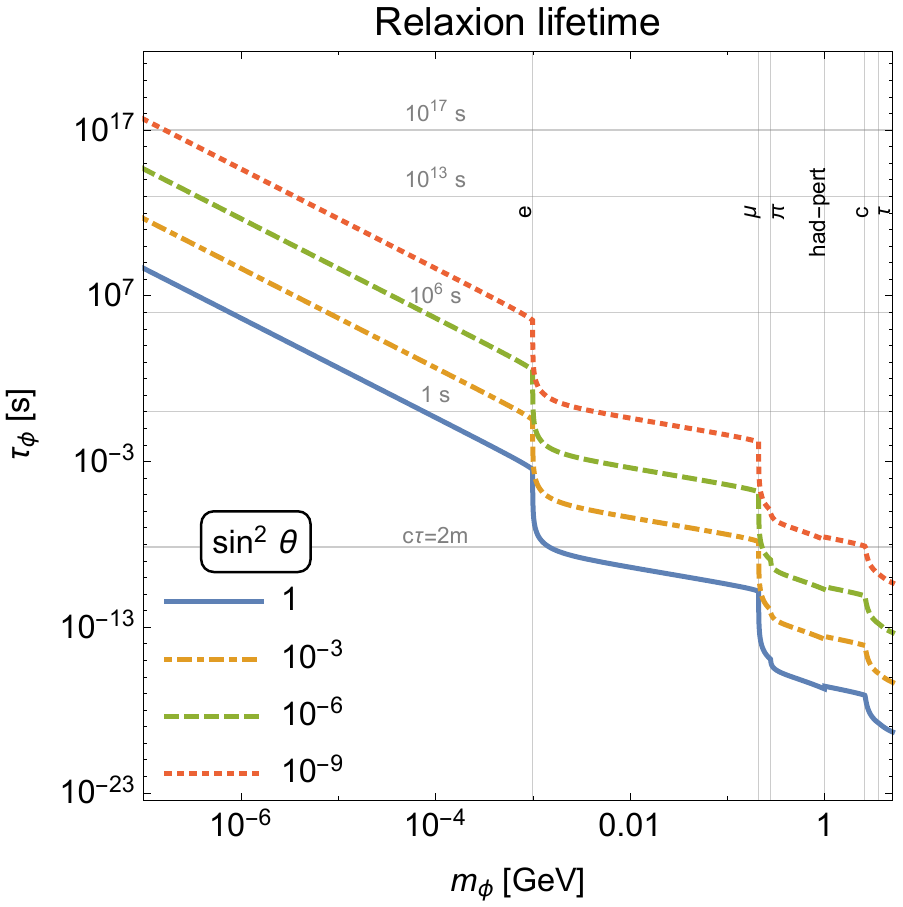} \label{fig:ctau}}}
\vspace{-2mm}
 \caption{Relaxion properties: (a) The relaxion mass $\mphi$ as a function of $\lbr$ for different values of $f$, where the vertical gray line indicates $\lbr^{*}$ that maximizes $\mphi$ for each $f$. Here $\lbrmax$  is the upper bound on $\lbr$ arising from the requirement of a non-tachyonic $\phi$ in \eq{tach} for $ \sin (\phi_0/f)=1/\sqrt{2}$.
 (b) The lifetime $\tau_{\phi}$ also depending on $\mphi$ and $\st$ with thresholds (vertical gray lines) and example values of $\tau_{\phi}$ (horizontal gray lines). The lifetime for any other $\st$ value can be obtained from the $\st=1$ line using $\tau_\phi\sim 1/\st$.}
 \end{figure}

In this and the next section we discuss in detail the bounds and the future probes for  relaxion-Higgs mixing, distinguishing between laboratory experiments, discussed here, and cosmological and astrophysical probes considered in section~\ref{cosmosec}.
As we will show below, the relaxion mass can range from  far below the eV-scale   to almost the weak scale. Therefore a variety of experiments is needed to look for the relaxion.
As the couplings to SM particles  are proportional to $\sin \theta$, a convenient plane to present the constraints is the $\st$-$\mphi$ plane. Before going into the details of the various constraints to be presented in figures~\ref{ev}--\ref{kev}, let us first identify the regions of the $\st$-$\mphi$ plane that are  relevant for relaxion models. For the convenience of the reader we repeat  the expressions for the mass and mixing angle of the relaxion in the small-mixing approximation from \eq{eq:sintj} and \eq{massmixj} substituting for definiteness $\sin (\phi_0/f)=\cos (\phi_0/f)=1/\sqrt{2}$,\footnote{Note that we have assumed $v^2_H\simeq v^2$ which amounts to ignoring, at most,  ${\cal O}(10 \%)$ effects (see section~\ref{secbr}).}
\begin{eqnarray}
\sin \theta &\simeq& j\frac{\lbr^4}{\sqrt{2}v f  m_h^2} \nonumber,\\
m_\phi^2 &\simeq& \frac{\lbr^4}{ 2f^2}\left(\sqrt{2}-\frac{j^2\lbr^4}{v^2 m_h^2}\right)=\frac{\lbr^4}{ \sqrt{2}f^2}\left(1-\left(\frac{\lbr}{{\lbrmax}}\right)^4\right)\, ,
\label{exprep}
\end{eqnarray}
\sloppy{where  $(\lbrmax)^2=2^{1/4} m_h v/j$ is the maximal allowed value of $\lbr$ that follows from Eq.~(\ref{tach}).}\footnote{It was shown in ref.~\cite{Choi:2016luu} that in $j=2$ models it is possible to have $\lbr\gtrsim m_h v$ with smaller  then $\mathcal{O}(1)$ values for $\sin (\phi_0/f)\lesssim v^2/\lbr^2$,  such that Eq.~(\ref{tach}) is still satisfied. In this work  we take $\mathcal{O}(1)$ values of $\sin (\phi_0/f)$, and in accordance with Eq.~(\ref{tach}), $\lbr\lesssim m_h v$ thus not considering this region of the parameter space. As the backreaction scale, $\lbr$, is in any case constrained to be less than a few times the weak scale (see section~\ref{secbr}), this is actually a narrow region of the parameter space where the constraints are expected to be  similar to those we obtain for $\lbr^2\sim m_h v$.} Other $\mathcal{O}(1)$ choices of $\phi_0/f$ lead to\enlargethispage{\baselineskip} slightly modified numerical values for $\mphi$, $\sin \theta $ and $\lbrmax$. These can be inverted to obtain
\begin{align}
 \lbr^2(\mphi, \sin \theta) &= \frac{2^{1/4}v m_h^2}{j}{\frac{\sin \theta}{\sqrt{\mphi^2+m_h^2\st}}}\,,\nonumber\\
 f(\mphi, \sin \theta) &= \frac{m_h^2 v \sin \theta}{j (\mphi^2+m_h^2 \st)}\,.\label{inv}
\end{align}
We use the equations above to make contours of constant $\lbr$ and $f$ in the $\st$-$\mphi$ plane in figure~\ref{ev},~\ref{mev} and~\ref{kev}. Although we have made the contours for the $j=2$ case, using \eq{inv} one can easily translate to the  $j=1$ case  by substituting $f\to2 f, \lbr \to \sqrt{2} \lbr$.  For relaxions heavier than 5\,GeV, the mixing can be ${\cal O}(1)$ and \eq{exprep} and \eq{inv} are no longer valid. Thus in section~\ref{lep} and figure~\ref{gev} where we consider relaxions in this mass range, we exactly diagonalize the mass matrices in appendix~\ref{app1} and~\ref{app2} to obtain the $\lbr$ and $f$ contours.

We see from   \eq{exprep} and   that if $\Lambda_{\rm br}$ is much smaller than $\lbrmax$, for a given $f$,  both $\mphi$ and  $\st$ increase with $\lbr$  and  we get $\st \sim \mphi^4$. This implies that in this regime, a light relaxion has typically a suppressed mixing.
However, if we take values of  $ \Lambda_{\rm br}$ close to $\lbrmax$,  this tendency does not hold anymore. This behaviour can be seen in figure~\ref{fig:mass} where we plot the relaxion mass as a function of  $\lbr$ for different values of $f$ taking $j=2$. We see that, for all $f$, the relaxion mass is maximum for $\lbr=\lbr^*=2^{-1/4}\sqrt{m_h v/j}$, and for larger values of $\lbr$  the relaxion mass drops rapidly with $\lbr$ as the term within  the parenthesis in \eq{exprep} becomes smaller. The relaxion mass can, in fact,  be made arbitrarily small by choosing a $\lbr$ that is  sufficiently close to its maximal value $\lbrmax$, while hardly changing $\st$.  In the $\st$-$\mphi$ plane this can be seen from the shape of the  $f$ contours in figure~\ref{ev} (and subsequent figures) for which two branches can be clearly identified. The region $\lbr>\lbr^*$  corresponds to  the  top left part of figure~\ref{ev}   where the  $f$ contours become nearly horizontal as $\st$ hardly changes but the mass can become arbitrarily small. The thick grey line in figure~\ref{ev}  is the contour $\Lambda_{\rm br}=0.99 \lbrmax$. The whole region above the this line, which we refer to  as the ``tuned region'', corresponds to the narrow region in the theory space  $0.99 \lbrmax< \Lambda_{\rm br}< \lbrmax$, marked  by the  thick black line in figure~\ref{fig:mass}. Therefore, in the following we will mostly discuss the ``untuned region"  $\lbr<\lbrmax$, which translates to
\begin{eqnarray}
s_\theta < 0.04~\frac{ m_{\phi}}{1 \gev}\,
\label{sthetaH}
\end{eqnarray}
and implies that in most  of the theoretical parameter space,  if we make the relaxion lighter, it also becomes more weakly coupled. We would like to point out that this is a general feature of Higgs portal models. For instance, consider the  potential~\cite{pospelov},
 \begin{eqnarray}
 V_{hp}=\frac{\hat{m}^2_{\phi}}{2} \phi'^2 +\frac{\hat{m}^2_h}{2} h'^2+ x \hat{m}_{\phi}  v h' \phi'
 \end{eqnarray}
where $\phi'$ and $h'$ are as defined in \eq{gen} while $\hat{m}^2_{\phi},\hat{m}^2_h$ and $x$  parametrise the couplings in a general Higgs-portal model.   For small mixing angles we get $\hat{m}_h=m_h$, the observed Higgs mass, and the condition $x<x_{\rm max}=m_h/v$ to ensure that the lighter eigenstate does not become  tachyonic. The region above the grey line in this case corresponds to  the small range $0.98~x_{\rm max}<x<x_{\rm max}$.

 The second restriction on the $\st-\mphi$ parameter space arises from the fact that we consider only the range,
\begin{eqnarray}
m_h<&f&<\mpl \nonumber\\
  \Rightarrow 6.5\times 10^{-5} \left(\frac{ m_{\phi}}{1~\gev}\right)^2>&\sin \theta& > 10^{-18} \left(\frac{ m_{\phi}}{1 {\rm~\mu eV}}\right)^2.
 \label{fregion}
\end{eqnarray}
The lower bound on $f$ arises from the fact that in our analysis of relaxion-Higgs mixing  we ignored any new states (for instance radial modes) that must exist below the scale $\Lambda=4 \pi f$ to UV-complete the backreacting sector. Thus our analysis holds only if both the Higgs boson and the relaxion have a mass much smaller than the mass scale of these UV states, i.e.~for $f\gtrsim m_h$.
 We will call the region defined by \eq{sthetaH} and \eq{fregion}  the `relaxion parameter space', i.e the region in the $\st-\mphi$ relevant for relaxion models.

 Let us now discuss the mass range the relaxion can have given these restrictions.  In the untuned region the relaxion can be made lighter either by decreasing $\lbr$ or increasing  $f$.  In our analysis we  do not consider  $f> \mpl=2\cdot 10^{18}\gev$, but as there is no strict lower bound on  $\lbr$, the relaxion can be made as light as we want by taking  sufficiently  small values of   $\Lambda_{\rm br}$. As discussed in section~\ref{intro}, however, lower values of  $ \Lambda_{\rm br}$ are theoretically disfavoured.
For instance if we require relaxion field excursions to be subplanckian this puts a bound $\st\lesssim10^{-27}$ as shown in figure~\ref{ev}. In the untuned region this can be translated to $\mphi  >0.001$\,eV\@. As the requirement of  subplanckian field excursions depends on quantum gravity assumptions and can be possibly evaded by UV model building, we will not take this as a strict bound and extend our constraints also to the  transplanckian region.

We now turn to the question of how heavy the relaxion can be. The  largest relaxion mass is  obtained for  the minimal value, $f=m_h$,   and weak scale values of $\lbr$ where the small-mixing approximation in \eq{exprep}  no longer holds.  In section~\ref{lep}, by exactly diagonalising  the mass matrices {in appendix~\ref{app1} and~\ref{app2}},  we find an upper bound  $\mphi\lesssim 60$\,GeV (see figure~\ref{gev}).

For readers interested in general Higgs portal models our analysis provides the complete constraints  in the untuned region of their parameter space apart from the area outside the region defined by \eq{fregion}. Whereas for $f>\mpl$, the constraints in the untuned part of the region   arise only from fifth force experiments and have been discussed elsewhere (see for instance~\cite{pospelov,Graham:2015ifn}), the region corresponding to $f\lesssim m_h$  can be potentially constrained only by some cosmological probes that we will mention in the next section but not fully~derive.

Before going into the details of the different experimental probes, a comment is in order. In the following we are going to study the constraints on the relaxion parameter space driven by its mixing with the Higgs.  As it is impossible to include the effects of the pseudoscalar couplings of the relaxion in a model-independent way we do not consider these. In any case, in existing explicit models, these couplings are generally not larger than the Higgs portal couplings as discussed in appendix~\ref{ps}.  An exception is the  pseudoscalar coupling to photons which can in some backreaction models (see appendix~\ref{ps}) be larger than the one induced via Higgs mixing.  In section~\ref{sect:summaryplots} we qualitatively comment on how our constraints would change if a large pseudoscalar coupling to photons is present.

In the following we describe the constraints on the relaxion in different mass ranges as the relaxion mass spans a wide  range from sub-eV values  to tens of GeV\@. While  relaxions heavier than a MeV can  be potentially probed by collider searches, the only laboratory probes  for sub-MeV relaxions are fifth force experiments. We  discuss these two categories separately starting with sub-MeV relaxions.

\subsection{The sub-MeV mass range}
\begin{figure}
\centerline{ \includegraphics[width=0.65\textwidth]{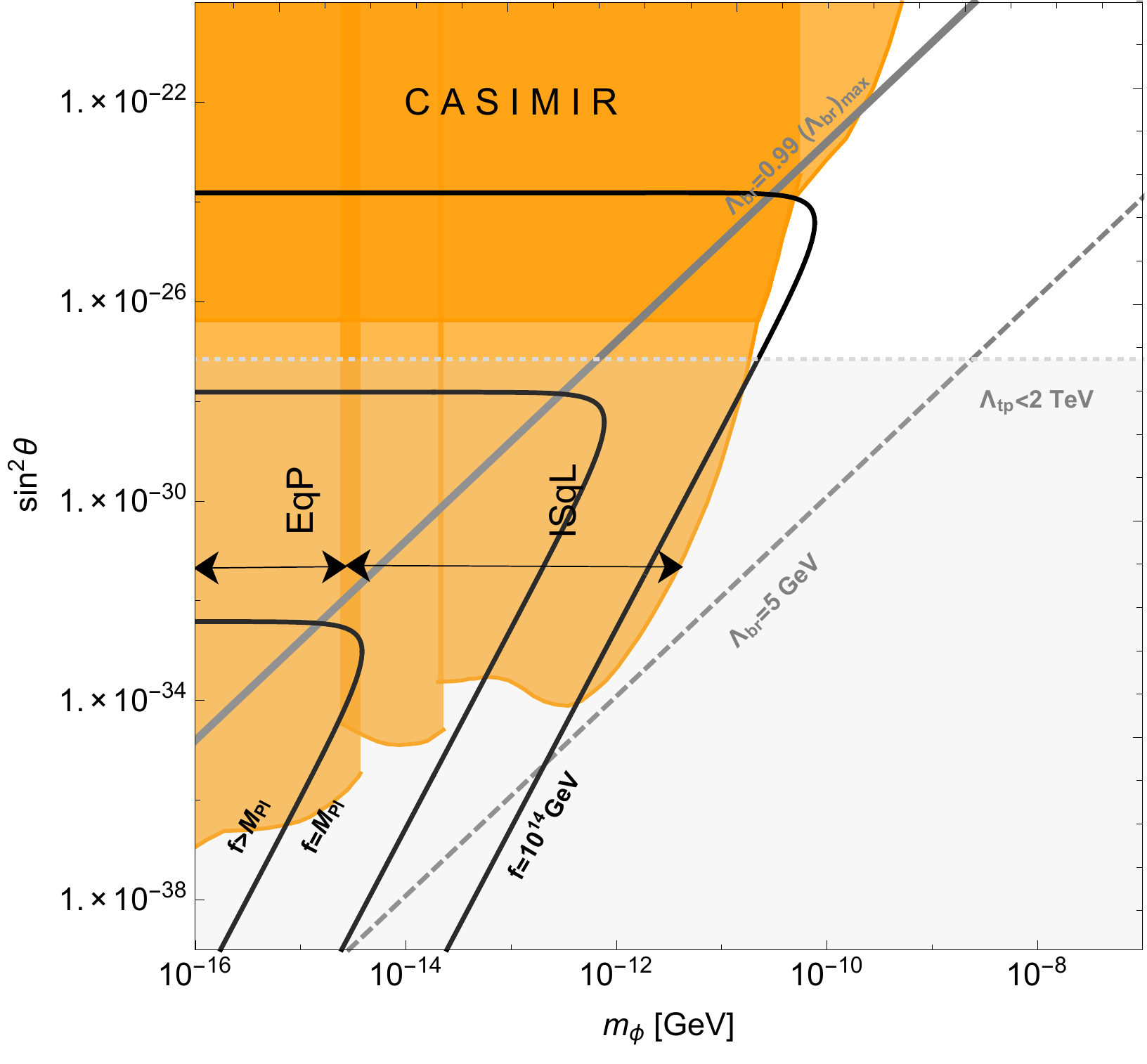}}
\vspace{-3mm}
 \caption{Constraints on the relaxion-Higgs mixing $\st$ for light relaxions with $\mphi$ between $10^{-16}\gev$ and $10^{-7}\gev$. Fifth-force experiments (orange) probe the lightest mass range via the equivalence principle (labelled as EqP), the inverse square law (ISqL) and the Casimir effect (Casimir).
 Contours of constant $\Lambda_{\rm br}$ (gray) for $\Lambda_{\rm br}=0.99\lbrmax\simeq 104\gev$ (gray, thick, solid)
 and $\Lambda_{\rm br}=5\gev$ (gray, dashed).   Here $\lbrmax$  is the upper bound on $\lbr$ arising from the requirement of a non-tachyonic $\phi$ in \eq{tach} for $ \sin (\phi_0/f)=1/\sqrt{2}$.
 Contours of constant $f=\mpl, 10^{16}\gev, 10^{14}\gev$ (black, solid).
 The light gray region below the dotted gray line corresponds to trans-Planckian field excursions $\Delta\phi>\mpl$ for $\Lambda=2\tev$.}
 \label{ev}
 \end{figure}

In this mass range  the relaxion has a  very large decay length making it impossible for collider searches to probe visible decays of the relaxion. This can be seen from figure~\ref{fig:ctau} where  we plot, using the expressions in appendix~\ref{appw},  the relaxion lifetime as a function of its mass for different choices of $\sin \theta$. \eq{sthetaH} implies for the considered mass range $\sin \theta\lesssim 10^{-9}$, and figure~\ref{fig:ctau} shows the corresponding enormous  rest frame decay length of $c\tau \gtrsim 10^{14}$ m. Therefore the only possible laboratory probes are either fifth force experiments, or  experiments looking for invisible particles.
This last class of experiments, at least at the moment, is not sensitive enough to provide constraints on the very small Higgs-relaxion mixing in this mass range~\cite{Harnik:2012ni}.
Fifth force experiments denote experiments which can detect the existence of a new degree of freedom by the corresponding new Yukawa-like force induced between two electrically neutral test bodies. A relaxion induces a spin-independent Yukawa force between two test bodies $A$ and $B$, defined by the potential
\begin{equation}
V= -G\frac{ m_A m_B}{r} \alpha_A \alpha_Be^{-r \; m_{\phi}}\,,
\end{equation}
where $m_A$, $m_B$ are their respective masses and  $\alpha_A $, $\alpha_B$ parametrise the   couplings of the relaxion to the two bodies. In Higgs portal models, the couplings are given by~\cite{pospelov}
\begin{eqnarray}
\alpha_A= \alpha_B= g_{hNN}\frac{\sqrt{2} \mpl}{m_{nuc}}s_\theta
\end{eqnarray}
where $g_{hNN} \simeq 10^{-3}$ and $m_{nuc}=1$\,GeV\@.
The sensitivity of the various fifth force experiments depends on the interaction length $\lambda$ which is related to the mediator mass $\mphi$~via
\begin{equation}
\lambda= m_{\phi}^{-1}= 1 \;  \mu m \frac{0.2\ev}{m_{\phi}}\,.
\end{equation}
Let us start discussing probes of new long-range forces going down from macroscopic length scales  to the pm scale of MeV particles. We present the bounds arising from these probes in figure~\ref{ev}. For very low masses  (below $3\cdot10^{-15}\gev$), the strongest constraint comes  from  the E\"ot-Wash experiments~\cite{Smith:1999cr,Schlamminger:2007ht} that looked for deviations from Einstein's weak equivalence principle (labelled as  EqP in figure~\ref{ev}) by precision measurements of the long-range force between a heavy attractor and two different test bodies in a torsion balance.
Let us notice that this experiment is able to constrain the Higgs portal down to very small couplings, but for masses lighter than $ 10^{-16}\gev$ the probed parameter space belongs to the tuned region (for other potentially relevant discussion of cosmological and/or low energy probes see for instance~\cite{Bellazzini:2015wva,Delaunay:2016brc}). Therefore, in figure~\ref{ev} we do not show relaxion masses of $\mphi<10^{-16}\gev$ although the EqP bound extends even further.
On shorter length scales, the mass range $3\cdot10^{-15}$-$10^{-11}\gev$, the strongest bounds arise  from constraints on violations of the inverse square law (labelled as InvSqL) that have been obtained by various experimental groups~\cite{Spero:1980zz,Hoskins:1985tn,Chiaverini:2002cb,Hoyle:2004cw,Smullin:2005iv,Kapner:2006si}.  The excluded region shown in figure~\ref{ev} is an envelope that contains bounds from all these experiments with the strongest one coming from the Irvine experiment in the mass range $3\cdot10^{-15}$-$5\cdot10^{-14}$\gev~\cite{Spero:1980zz,Hoskins:1985tn}, from the E\"ot-Wash 2006 experiments  in the mass range $5\cdot10^{-14}$- $2\cdot10^{-12}\gev$~\cite{Kapner:2006si} and from the Stanford experiment~\cite{Chiaverini:2002cb,Smullin:2005iv} in the mass range $2\cdot10^{-12}$- $5\cdot10^{-11}\gev$.
Finally we also show the constraints from tests of the Casimir force~\cite{Bordag:2001qi,bordag2009advances}, the force induced by the zero point energy of the electromagnetic field when two conductors are brought very close to each other.  While these bounds from the tests of the Casimir effect are weaker than the bounds of the torsion balance experiments below $10^{-11}\gev$, they are the strongest bounds above this mass as shown in figure~\ref{ev}.
The shaded area below the horizontal light gray, dotted line ($\st\leq 10^{-27}$) shows  the region where  the relaxion has transplanckian excursions  for any value of the cut-off scale  $\Lambda>2~\tev$ (see \eq{trp}).

For heavier particles, i.e.~shorter-range forces, the sensitivity is even lower. The intermediate region, between  10\,eV and 1\,MeV, is the most challenging region to probe in laboratories. The most sensitive experiment in this mass region are  neutron scattering experiments that  test the existence of a new sub-MeV boson  based on their influence on the neutron-nucleus interaction. These experiments set a  very weak bound, $s_\theta \lesssim 0.1$~\cite{Nesvizhevsky:2007by,Pokotilovski:2006up}, and are therefore incapable of probing a relevant   region of the parameter space. In a subset of this mass range from a keV to an MeV (shown later in figure~\ref{kev}) the relaxion parameter space can be probed only by astrophysical and cosmololgical observations  to be discussed  in detail in the next section. The 10\,eV-keV mass range, on the other hand, is largely unconstrained as shown in figure~\ref{ev}.

Let us conclude this subsection by commenting  that fifth force experiments  are a unique probe of light states like relaxions that couple to electrons and nucleons as CP-even scalars. Axions, for instance, do not give rise to spin-independent long range forces at leading order  because of their pseudoscalar nature and are thus only weakly constrained by fifth force experiments. Therefore, different laboratory probes have been proposed to circumvent this problem.
This is the case for light shining through the wall (LSW) experiments~\cite{Redondo:2010dp}, which  are also sensitive to Higgs portal models~\cite{Ahlers:2008qc}.
However, their reach is  too limited to compete with fifth force experiments  and therefore these do not appear in our plot.

\subsection{Relaxion masses between the MeV- and the weak scale}
Let us now study the region  of parameter space where the relaxion mass is above the electron threshold and thus it can decay into SM fermions.
Furthermore, as shown in figure~\ref{fig:ctau}, in this region the relaxion  has a shorter lifetime and can be directly searched for in laboratory facilities.
Let us further distinguish two sub-regions based on the different relevant probes.
The bounds in the MeV-5\,GeV  mass range are presented in figure~\ref{mev}, including also astrophysical and cosmological constraints which will be discussed in the next section. Figure~\ref{gev} presents the bounds in the GeV region.

\subsubsection{The 1 MeV--5 GeV range}
\label{lep}
This region of the parameter space is well covered by  rare $K$- and $B$-meson decays at proton beam dump and flavour experiments.
Crucial for both kinds of experiments is the possibility of producing a relaxion in rare decays of $K$- and $B$-mesons. In flavour experiments that probe rare decays,  constraints are put on the  branching ratios~\cite{Clarke:2013aya}
\begin{align}
\br( K\rightarrow \pi^+ \phi)& =  0.002 \st \; \frac{ 2 |  p_{\phi}|}{ m_K}  \, ,\\
\br( B\rightarrow K^+ \phi)& =     0.5 \st \;  \frac{ 2 |  p_{\phi}|}{ m_B} {\cal F} _K^2( m_{\phi}) \, ,
\end{align}
where $  p_{\phi}$ is found using two-body kinematics and ${\cal F}_K$ is defined in~\cite{Clarke:2013aya}.
Even  in proton beam dump experiments, rare mesons decays are the main the production mode  of the relaxion.
The smallness of the branching ratio  is overcome by the large luminosity. Electron beam dump experiments do not have any sensitivity to Higgs-relaxion mixing due to the suppressed   electron Yukawa coupling.

\begin{figure}
\centerline{ \includegraphics[width=0.65\textwidth]{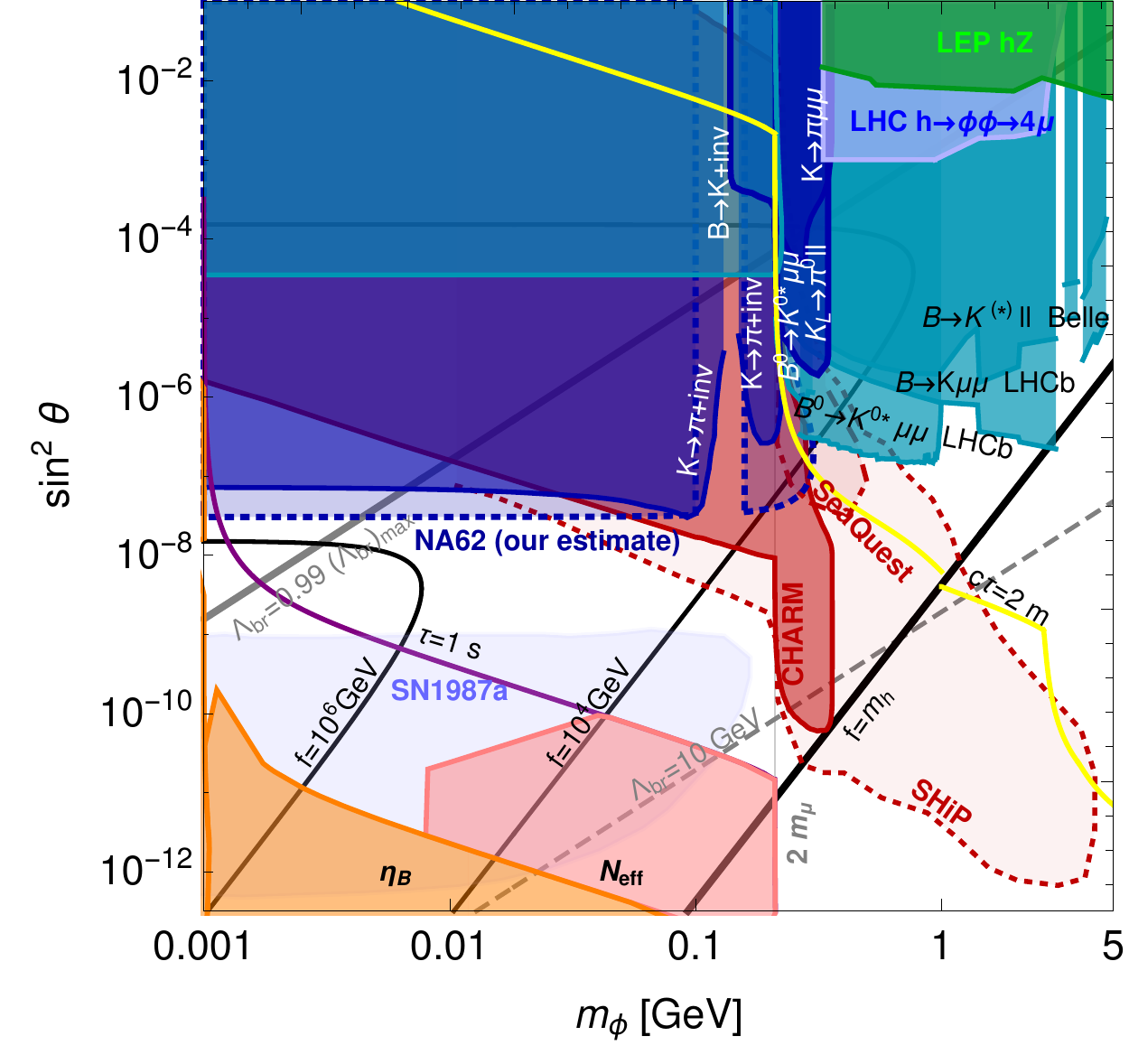}}
\vspace{-3mm}
 \caption{Constraints on the relaxion-Higgs mixing $\st$ for relaxions with $\mphi$ between MeV and 5\,GeV\@.
 The laboratory probes include:   proton beam dump experiments (red for CHARM, light red for the projected sensitivity for SHIP and SeaQuest),  $K$-meson decays (blue,  our conservative projection from NA62 in a lighter shade of blue), $B$-meson decays (turquoise), LHC search for $ h\rightarrow 4 \mu$ (light blue) and LEP (green).
 Astrophysical and cosmological  probes include the Supernova 1987a (pale violet, labelled as SN), $ \eta_b$ (orange) and $N_{\text{eff}}$( pink).
 Contours  for $\Lambda_{\rm br}=0.99\lbrmax\simeq 104\gev$ (gray, thick, solid), $\Lambda_{\rm br}= 10 \gev$ (gray, dashed),
  $f/\gev=10^6,  10^4, 125$ (black, solid) are presented. Here $\lbrmax$  is the upper bound on $\lbr$ arising from the requirement of a non-tachyonic $\phi$ in \eq{tach} for $ \sin (\phi_0/f)=1/\sqrt{2}$.
  The vertical light gray line corresponds to the contour for the relaxion mass at the muon threshold; the yellow contour corresponds to $c \tau =2$\,m and the purple one to $ \tau = 1$\,s.
 }
 \label{mev}
 \end{figure}

\paragraph{Beam dump experiments:} In proton fixed target experiments,  relaxion beams are produced from meson decays and constraints are imposed by looking at its visible decays.  The region to the left of the $c \tau=2$\,m line in figure~\ref{mev} is roughly the region where the relaxion decay length in the lab frame is greater than about 100\,m (assuming a relativistic boost factor $\gamma \beta \sim 50$~\cite{Clarke:2013aya}). Thus this is the region relevant for beam dump experiments looking for long lived particles.
We will discuss here the sensitivity of  the CHARM experiment and future experiments such as SHIP~\cite{Alekhin:2015byh} and SeaQuest~\cite{Gardner:2015wea}. NA62 is also planning a beam dump run  as proposed in ref.~\cite{Dobrich:2015jyk}.

The CHARM beam dump experiment performed a search for long-lived axion-like particles decaying to $e^+e^-$, $\mu^+\mu^-$ or $\gamma\gamma$  in collisions of a $400\gev$ proton beam on a copper target~\cite{Bergsma:1985qz}
 with a 35\,m long detector located 480\,m from the target.
 This search can be also reinterpreted in  the context of Higgs portal models~\cite{Bezrukov:2009yw,Clarke:2013aya,Schmidt-Hoberg:2013hba,Alekhin:2015byh}, where the scalar $\phi$ is predominantly produced in rare decays of $K$- and $B$-mesons.  In such an experiment around $10^{17}$ kaons  and  $10^{10}$ $B$-mesons are produced per year. Figure~\ref{mev} shows that CHARM (dark red) is able to constrain only masses below the kaon threshold. The limited sensitivity is due to the lower $B$-meson luminosity and the large distance between target and detector.
  The large distance, on the other hand, is good to probe the low mass region where the relaxion has a longer decay length as shown in figure~\ref{fig:ctau}. In figure~\ref{mev} we show  also the projections (in lighter red) for several future proton beam dump experiments such as SHIP (dotted) and SeaQuest (dash-dotted). While the reach of NuCal exceeds CHARM for a scalar/pseudoscalar with couplings only to photons~\cite{Dobrich:2015jyk}, the NuCal bound in the presence of Yukawa-like couplings to fermions is in ref.~\cite{Dolan:2014ska} found to be weaker than the  CHARM limit and therefore we omit NuCal in figure~\ref{fig:ctau}.
 If in  future beam dump experiments  the detector is  closer to the target than  in the case of CHARM, good improvements over CHARM can be achieved for relaxion masses heavier than the muon threshold where the lifetime is shorter.
 As already noticed, lighter relaxions have a longer lifetime and therefore the CHARM bound in this region can be improved at proton fixed target experiments by looking for invisible new particles.
 The present sensitivity, however, is limited in the region of the parameter space relevant to our scenario~\cite{Batell:2009di,Morrissey:2014yma,Frugiuele:2017zvx,Coloma:2015pih}.


\paragraph{Rare meson decays:} Rare decays of $K$-, $B$- and $\Upsilon$-mesons can be mediated by a light scalar particle $\phi$, and therefore bounds on their branching ratios constrain the relaxion-Higgs mixing angle.
In figure~\ref{mev} the turquoise region corresponds to the bounds on $B$-decays and the blue region to $K$-decays.
We do not show  bounds coming from rare $\Upsilon$ decays since they are always weaker than other existing bounds.

Let us first discuss how $B$-decays constrain the relaxion-Higgs mixing.
Both Belle and LHCb are sensitive to the decay process of $B^\pm \rightarrow K^\pm \phi \rightarrow K^\pm l^+l^-$ with $l=\mu$ at LHCb and $l =\mu,~e$ at Belle~\cite{Aaij:2012vr,Wei:2009zv}.
In the experimental analyses, the regions of $m_{ll}$ in $[2.95\gev,~3.18\gev]$ and in $[3.59,~3.77\gev]$ are vetoed in order to suppress the background from the $J/\psi$ and the $\psi'$ resonances, respectively.
Figure~\ref{mev} shows the constraints on $\st$ derived in~\cite{Schmidt-Hoberg:2013hba} using the upper  bound  on the branching ratio as a function of the dilepton invariant mass provided by  LHCb  and Belle (both in turquoise).  The figure indicates an almost comparable sensitivity of both experiments in the mass region above $ \sim 300$\,MeV.\footnote{As LHCb places slightly stronger constraints, we will omit the Belle results in the summary plots in section~\ref{sect:summaryplots}.} In addition, LHCb constrains
$\br(B^0 \rightarrow K^{0*} \phi)\,\br(\phi\rightarrow \mu^+\mu^-)$
as a function of the mass and the lifetime of a new boson $\phi$. Figure~\ref{mev} includes this LHCb search as a bound on $(\mphi,\,\st)$ for $\mphi\leq 1\gev$ as presented in ref.~\cite{Aaij:2015tna} for a scalar mixing with the Higgs from the model of ref.~\cite{Bezrukov:2014nza}, which also applies to the relaxion case. It improves the previous LHCb bound by appropriately an order of magnitude.
Using the full 2-dimensional information provided in ref.~\cite{Aaij:2015tna}, an extension of the bound up to $\mphi\leq 4.35\gev$ would be possible.

In the range of $0.212\gev\leq \mphi \leq 0.3\gev$~\cite{Hyun:2010an}, Belle has performed a dedicated study of $B^0\rightarrow K^{*0} \mu^+ \mu^-$, searching for a peak in the dimuon spectrum in order to enhance the sensitivity; this bound is also shown in figure~\ref{mev}.
For even lighter masses, the limit on the  $B \rightarrow K +\textrm{invisible}$ from Belle and BaBar (also indicated in turquoise) constrains the relaxion parameter space in a region where the relaxion has a long decay length. In both of the above cases we have used the constraints derived by ref.~\cite{Clarke:2013aya}.

Let us now discuss the constraints set  by the searches for visible and invisible rare $K$-meson decays, using again the results of ref.~\cite{Clarke:2013aya}. These are shown in dark blue in figure~\ref{mev}.
A search for $K_L \rightarrow \pi^0 l^+l^-$ has been performed at KTeV/E799~\cite{AlaviHarati:2000hs,AlaviHarati:2003mr} and translated into bounds on a pseudoscalar~\cite{Dolan:2014ska} and scalar~\cite{Alekhin:2015byh} mediator of this decay. The corresponding constraint (dark blue) in figure~\ref{mev} is stronger above the muon threshold, where it surpasses the current constraints from $B$-decays, and much weaker when the only visible decay mode is into electrons (shown as a dark blue line).
The branching ratio of $K^\pm \rightarrow \pi^\pm \mu^+\mu^-$ has been measured by the NA48/2 fixed target experiment~\cite{Batley:2011zz} at the CERN SPS. Despite the good agreement of the branching ratio with the SM prediction, the resulting bound on $\st$ is weaker than those derived from the $B$-decays into visible final states due to the relative CKM suppression of $V_{ts}\cdot V_{td}$ compared to $V_{tb}\cdot V_{ts}$ in the $W-t$ - loop of the penguin diagram.

\enlargethispage{\baselineskip}

On the other hand, strong constraints can be set looking at invisible $K$-decays and this is a promising search for light relaxions.
Indeed for small enough couplings  and/or light enough masses -- more precisely the region to the left of the $c\tau=2$\,m contour in figure~\ref{mev} --   the relaxion decays outside the detector (see also figure~\ref{fig:ctau}).  Searches for the invisible $K$-decay $K^\pm \rightarrow \pi^\pm +\textrm{invisible}$  have been  performed  by the E787~\cite{Adler:2004hp} and the E949~\cite{Artamonov:2009sz} experiments at BNL, also considering two-body decays and providing limits on $\br( K\rightarrow \pi^+ \phi)\,\br(\phi \to \textrm{invisible})$ as a function of $\mphi$. The constraints on the Higgs portal model coming from these analyses were previously studied in~\cite{Clarke:2013aya},
which, however, focussed only on the region above 100\,MeV,
while the search is sensitive also to lighter relaxions.
Therefore, we extended the analysis to lower masses as shown in figure~\ref{mev}; the gap in the mass range  $0.1\gev \leq \mphi \leq 0.21\gev$  is due to the fact that the region around $m_{\pi^\pm}$ has been vetoed.
 We find that the E949 experiment gives stronger  limits than  the CHARM beam dump experiment for relaxions with $\mphi\leq10\mev$. Indeed, lighter relaxions have a larger decay length, thus they are most likely detected as invisible particles.
Therefore, this is one of the most promising regions for rare $K$-decay measurements to probe new physics.
For instance the CERN experiment NA62  will improve the present limit on invisible $K$-decays by almost an order of magnitude. They expect to see 90 SM signal events and 20 background events  in two years~\cite{Proceedings:2013iga}. Using only this  information about the total rate and no information about the differential distribution of the SM and background events, we show a conservative estimate of the 95$\%$ CL excluded region  in light blue in figure~\ref{mev} where we have assumed a 10$\%$ theoretical error~\cite{manos}. The gap in the excluded region is again  due to the veto around the charged pion mass, $100\mev\lesssim m_\phi\lesssim 160\mev$~\cite{Proceedings:2013iga}.

Finally, for GeV-scale masses we see from figure~\ref{mev} that some regions of the parameter space are bounded by LEP and LHC searches that we describe in detail in the next section.

\subsubsection[The \texorpdfstring{$\mphi> 5$}{m(phi)>5} GeV mass range]{The \texorpdfstring{\boldmath $\mphi> 5$}{m(phi)>5} GeV mass range}

\begin{figure}
\centerline{ \includegraphics[width=0.65\textwidth]{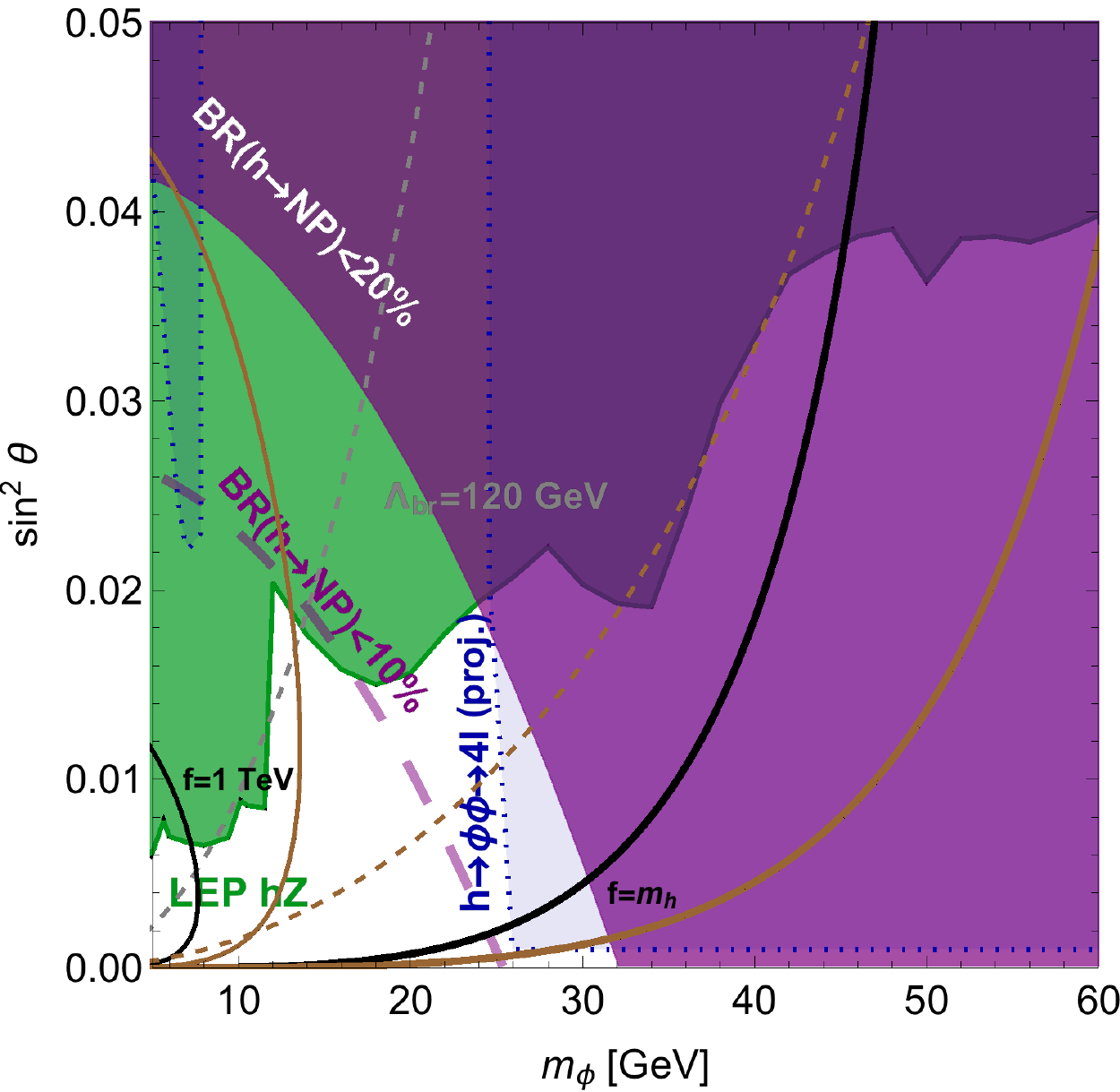}}
\vspace{-2mm}
 \caption{Constraints on the relaxion-Higgs mixing $\st$ for relaxions with $\mphi$ between $5\gev$ and $90\gev$ from LEP and the LHC: 4-fermion final states from Higgs strahlung at LEP (green, labelled as LEP hZ);  Higgs decays to NP with $\br(h \rightarrow\rm{NP})\leq20\%$ at the LHC (purple, solid) as well as a projection for $\br(h \rightarrow\rm{NP})\leq10\%$ (purple, dashed);
 explicit searches for $h\rightarrow\phi\phi$ with final states
 $4\tau$ (dark blue, dotted, $\mphi<10\gev$, Run 3 projection) and $2\mu2b$ (dark blue, dotted, $\mphi>25\gev$, Run 3 projection).
 Contours for $\lbr=120$\,GeV (gray, dashed for $j=2$; brown, dashed for $j=1$), $f=m_h$ and $f=1\tev$ (black for $j=2$, brown for $j=1$).
}
 \label{gev}
 \end{figure}

Finally we consider the mass region $\mphi>5\gev$ where the mixing angle $\sin \theta$ can become ${\cal O}(1)$ and the expressions in Eq.~(\ref{exprep}) do not apply anymore. To compute the mixing angle, $\sin \theta$, and the mass, $\mphi$, as functions of $\lbr$ and $f$, we therefore exactly diagonalise the mass matrix in appendices ~\ref{app1} and \ref{app2} for the $j=2$ ($j=1$) case. We fix the value of the unknown $\lambda$ by demanding that we obtain the observed Higgs mass for the heavier eigenvalue. This is how we obtain $\lbr$ and $f$ contours in figure~\ref{gev}. It is in this region that we obtain  lowest values of $f$ close to $m_h$.  As discussed in the beginning of this section, for even smaller values of $f<m_h$ our analysis of relaxion-Higgs mixing does not hold anymore.

\paragraph{LEP constraints:} In the high-mass range, LEP and the LHC provide useful constraints on the mass and coupling of the relaxion. At LEP, the Higgs-strahlung  process of, $e^+e^- \rightarrow Z \rightarrow Z^* h$ with $Z^*$ and $h$ each decaying to a pair of fermions, is sensitive to the Higgs-relaxion mixing. If $\mphi<2m_{\mu}$, $\phi$ escapes the detector. For visible decays above the dimuon threshold, L3~\cite{Acciarri:1996um} sets the most stringent bounds in the range $\mphi< 11.5\gev$ whereas for $12\gev\leq \mphi \leq 116\gev$ the combination of the four experiments ALEPH, DELPHI, L3 and OPAL at LEP~\cite{Schael:2006cr} constrains this process most strongly. The experiments provide a mass-dependent upper bound on the ratio of cross sections~\cite{Schael:2006cr},
\begin{equation}
 \mathcal{S}_{95} = \sigma_{\textrm{max}}/\sigma_{\textrm{SM}},
\end{equation}
where $\sigma_{\textrm{max}}$ is the largest cross section $\sigma(e^+e^- \rightarrow Z \rightarrow Z^* \phi)$ compatible within the $95\%$ CL with the combined data sets, and $\sigma_{\textrm{SM}}$ is the SM reference cross section $\sigma(e^+e^- \rightarrow Z \rightarrow Z^*\,H_{\textrm{SM}})$. In Higgs portal models, the ratio of $\phi Z$-production  to the SM Higgs production cross-section for the same mass is just $\st$, so that   $\mathcal{S}_{95} $ can be directly interpreted as the $95\%$ CL upper bound on $\st$.  We show the parameter space excluded by LEP in green, labelled by ``LEP hZ", in figure~\ref{gev}.

\paragraph{\boldmath Higgs coupling bound on ${h}\rightarrow \phi \phi$:} Finally we discuss how Higgs coupling measurements at the LHC constrain the $h \rightarrow \phi \phi$ process. The strongest constraint on the partial width to this non-standard decay channel arises from the potential dilution it can cause to the visible decay channels of the Higgs boson to SM particles. While such a dilution of the visible decay channels may be compensated by increased scaling factors of the couplings~\cite{Bechtle:2014ewa}, this is not the case in Higgs portal models (like the relaxion case we are considering) where the Higgs boson couplings are universally suppressed by $\cos\theta$ with respect to their SM values. This configuration with one universal coupling modifier and  non-standard decay channels has been considered in ref.~\cite{Bechtle:2014ewa}. Therefore we apply their upper limit on the  Higgs branching ratio to non-standard channels from a fit to the data of ATLAS and CMS at 8\,TeV with \texttt{HiggsSignals}~\cite{Bechtle:2013xfa}:
\begin{equation}
 \br(h \rightarrow \rm{NP}) \leq 20\% ~~\rm{at}~ 95\%~ CL\,.
\end{equation}
We compute the partial width of $h$ into $\phi\phi$,
\begin{equation}
 \Gamma(h\rightarrow \phi\phi) = \frac{1}{32\pi}\frac{|g_{h\phi\phi}|^2}{m_h}\sqrt{1-\frac{4\mphi^2 }{m_h^2 }}\,,
\end{equation}
using the coupling $g_{h\phi\phi}$ that has been derived exactly in Eq.~(\ref{eq:gmix}) in appendix~\ref{apphpp} for $j=2$, taking  $h-\phi$ mixing into account.  The $h \phi \phi$ coupling is parametrically different in $j=1$ models and has not been considered here (see appendix~\ref{apphpp}) . This allows to set bounds on the relaxion parameter space via
\begin{align}
 \br(h\rightarrow\textrm{NP} ) = \br(h\rightarrow \phi\phi) &= \frac{\Gamma(h\rightarrow \phi\phi)}{\Gamma(h\rightarrow \phi\phi)+\cos^2 \theta\Gamma_h^{\rm{SM} }} \stackrel{!}{\leq} 20\% \label{eq:BRinv}\,.
\end{align}
where  $\Gamma_h^{\rm{SM} }=4.12\mev$~\cite{Heinemeyer:2013tqa}.

\paragraph{Higgs decays to two relaxions at the LHC:} In addition, the explicit searches at the LHC for non-standard decays of the Higgs boson (see e.g.\ ref.~\cite{Curtin:2013fra}) with a mass of $m_h=125\gev$ include the decay channel of the Higgs boson into two lighter scalars (or pseudoscalars) $\phi$ that each further decay into a pair of fermions $f$ or photons $\gamma$:
$h \rightarrow \phi \phi \rightarrow 4f/4\gamma$ at ATLAS~\cite{Aad:2015oqa,Aad:2015bua} and CMS~\cite{CMS:2016cel,CMS:2015ooa,CMS:2016cqw,Khachatryan:2015nba}. Their results can be interpreted as bounds on the decay of the Higgs boson into two relaxions that further decay into the analysed final states. So far, data is only available from LHC Run 1 at $8\tev$. In ref.~\cite{CMS:2016cqw}, the CMS searches with the final states $4\mu, 4\tau, 2\mu2\tau$ and $2\mu2b$ have been translated into upper bounds on
\begin{equation}
R_{h\phi\mu}:=\sigma_h/\sigma_h^{\textrm{SM} } \times \textrm{BR}(h\rightarrow \phi\phi)\cdot \textrm{BR}(\phi\rightarrow\mu\mu)^2
\end{equation}
at the $95\%$ CL
under the assumption of
$g_{\phi f} \propto m_f$,
which holds also in the relaxion case, see section~\ref{secmix}. Therefore, the prediction of $R_{h\phi\mu}$ depending on $\mphi$ and $\st$ compared to the experimental limits provides bounds in the $(\mphi,~\st)$ plane for those values of $\mphi$ that are covered by the set of searches. The mass range of $0.25\,\gev\leq \mphi \leq 3.55\gev$ is covered by the $4\mu$ final state, and the current data is sufficient to exclude parts of the parameter space shown in figure~\ref{mev} (blue), but not stronger than the flavour bounds in this region.
At higher masses, the $2\mu2b$ final state is particularly sensitive due to the enhanced branching ratio of $\phi\rightarrow bb$ compared to $\phi\rightarrow\mu\mu$. However, the ATLAS and CMS searches based on the data from Run 1 do not constrain the relaxion parameter space beyond the constraints derived from the Higgs couplings fit. Assuming an improvement of the experimental limits by a factor of 10, which we very roughly estimate (neglecting the change of systematic uncertainties and not combining channels and experiments) as the reach during Run 3, we also show projections of the bounds for Run 3 (dark blue, dotted). While the $4\tau$ final state at the LHC will not set stronger bounds than Higgs-strahlung at LEP, the constraints coming from the $2\mu2b$ channel might provide an improvement comparable to the projected Higgs couplings fits.

To summarise, figure 4 visualises that the bounds from LEP and the LHC are comple-mentary in the sense that LEP is more constraining on $\st$ for $\mphi <$25\,GeV whereas the indirect constraint from the bound on the decay width into NP final states at the LHC sets a stronger constraint for $\mphi >$25\,GeV\@. Again we show contours of constant $\lbr$ and $f$ which, as we already mentioned,  have been obtained by exact diagonalisation of the mass matrices in appendices ~\ref{app1} and~\ref{app2}. We show the contours for $\lbr=120$\,GeV for $j=2$ (gray, dashed) and $j=1$ (brown, dashed), $f=m_h$ and $f=1\tev$ for both the $j=2$ (black) and the $j=1$ case (brown).

\section{Cosmological  and astrophysical probes of  relaxion-Higgs mixing}
\label{cosmosec}

As discussed in the previous section, laboratory measurements  can probe a significant region of the relaxion parameter space. However, in the sub-MeV region, before the fifth force experiments start to gain sensitivity in the sub-eV region, a large portion of the parameter space is left unconstrained. In this section we show how astrophysical and cosmological probes can explore  part of this region of the  parameter space, as shown in figure~\ref{kev}, and also provide relevant bounds if the relaxion mass is in the MeV-GeV range (also  shown in figure~\ref{mev}).  In order to identify the part of the parameter space most relevant for relaxion models and to gain an understanding of the theory contours in figure~\ref{kev}, we refer the reader to the discussion at  beginning of section~\ref{lab}.

\subsection{Cosmological probes}

Late relaxion decays can be constrained by a variety of cosmological probes such as light element abundances, CMB spectral distortions and distortions of the diffuse extragalactic  background light (EBL) spectrum. In this section we first compute the relaxion abundance generated by misalignment and thermal production and then use this result to study how these bounds apply to our scenario.

\subsubsection{Relaxion abundance}

\paragraph{Misalignment production:} During inflation the expectation value of the field $\phi$, $\langle \phi \rangle$,  satisfies the classical equation of motion. Quantum fluctuations lead to a spreading of the field around this classical value. The spreading is given by (see for instance ref.~\cite{bcn})
\begin{eqnarray}
\frac{d \Delta \phi^2}{dN_e}=\frac{H_I^2}{4 \pi^2}-\frac{2}{3 H_I^2}\left\langle\Delta \phi \frac{\partial V}{\partial \phi}\right \rangle\, ,
\end{eqnarray}
where $\Delta\phi=\phi-\langle\phi \rangle$ and $\Delta\phi^2=\langle(\phi-\langle\phi \rangle)^2\rangle$, $H_I$ is the Hubble scale during inflation and $N_e$ is the number of e-folds. We see that the spreading stops when the r.h.s.\  above vanishes, that is for
\begin{eqnarray}
\Delta \phi \simeq \frac{3 H_I^4}{8 \pi^2 V'(\phi)}  \lesssim\frac{3}{8 \pi^2 }\left(\frac{\Lambda^4_{\rm br}}{f}\right)^{1/3} \, ,
\label{pmax}
\end{eqnarray}
where $V'(\phi)=\partial V/\partial \phi$, and  to obtain the inequality we have used the  requirement that the dynamics of the relaxion is dominated by classical rolling and not quantum fluctuations, $H_I < (V'(\phi))^{1/3}$ (see ref.~\cite{kaplan}). This gives us the misalignment of $\phi$  from its classical value just after inflation.  After this, the Universe goes through a phase of radiation domination.  If the temperature of the Universe is below the temperature $T_0$ with
 \begin{eqnarray}
 H(T=T_0)=\frac{m_\phi}{3} ~~~
\Rightarrow~~~  T_0=\left(\frac{45}{4 \pi^3 g_*}\right)^{1/4}\sqrt{\frac{m_\phi}{3} \mpl}\,,
 \end{eqnarray}
the relaxion field oscillates around the minimum. This leads to an energy density, $\rho_{\phi m}$, and an effective non-relativistic number density, $n_{\phi m}$, given by
\begin{eqnarray}
\rho_{\phi m}&=&\frac{m_\phi^2 \Delta \phi^2} {2} \,,\\
n_{\phi m}&=&\rho_\phi/m_\phi\,,
\end{eqnarray}
and thus results in a comoving number density,
\begin{eqnarray}
Y_{m}=\frac{n_{\phi m}}{s} \lesssim   \frac{m_\phi  \; \Delta \phi_{\rm max}^2}{2 s}\,,
\label{mis}
\end{eqnarray}
where $\Delta \phi_{\rm max}$ is that maximal value of $\Delta \phi$ given by \eq{pmax},  the entropy density,  $s=0.44~g_*^S(T_i)\, T_i^3$ and $g_*^S(T_i)$ is the effective number of degrees of freedom in entropy at the temperature $T_i$. If the reheating temperature is larger than $T_0$, then $T_i=T_0$, otherwise  $T_i$
 is the reheating temperature.

\paragraph{Thermal production:} Relaxions can be thermally produced by the process $HH \to \phi \phi$ at temperatures above the Higgs mass, by the processes $q(g)+g \to q(g)+ \phi$  at  temperatures below the electroweak critical temperature, $T_{\rm EW}$,  by the pion-relaxion conversion process $N+\pi \to N+ \phi$ at temperatures below $\Lambda_{\rm QCD}$, and finally by inverse decays. Let us consider these processes one by one.

At temperatures above the electroweak critical temperature, $T_{\rm EW} \sim m_h v/m_t$  the Higgs portal mixing in \eq{massmixj}  is absent and the relaxion  interacts only with the Higgs doublet. The main production mode of the relaxion is then the process $HH \to \phi \phi$ via the coupling
\begin{eqnarray}
 g^2(H^\dagger H) \phi^2\,.
 \end{eqnarray}
 Note that any contribution to the process from the backreaction potential is absent,  because in both the non-perturbative axion and the peturbative familon model, the backreaction term dissolves at high temperatures. In the axion case, the potential becomes negligible at high temperatures because instanton effects become very weak as the non-abelian gauge  coupling becomes perturbative. In the familon model  the Coleman-Weinberg potential gets no contributions from  momenta above $m_{N^c}$ so that for $T\gtrsim m_{N^c}$ the backreaction potential vanishes also in this case. The comoving number density for $\phi$ resulting from this process has been computed in ref.~\cite{bcn} to be
\begin{eqnarray}
Y_{H^2}\simeq13.7~g^4  \frac{0.278}{g_* (T_{\rm EW})}\frac{M_{pl}}{T_{\rm EW}}\, ,
\label{H2}
\end{eqnarray}
where we have not considered any contribution above the electroweak critical temperature and  $g_* (T_{\rm EW})$ is the effective number of relativistic degrees of freedom in energy density at the temperature $T_{\rm EW}$.  As we shall see in the following, this is negligible compared to the production via the relaxion-Higgs mixing in the EW broken phase.

Now let us consider relaxion production in the EW broken phase, that is, production at temperatures much below the critical temperature of the electroweak phase transition, $T_{\rm EW}\sim m_h v/m_t=180\gev$. In order to ensure that any finite temperature effects are negligible, we take $T<T_0= 20\gev$ so that  we always have $(T/T_{\rm EW})^2 \ll 1$.  At these temperatures $t, h, Z, W^\pm$ are not relativistic and their densities are Boltzmann-suppressed.  We thus ignore any contribution to thermal production of relaxions from processes involving these states for $T\lesssim 20$\,GeV and ignore any contribution at all  from the temperature range $20~{\rm GeV}< T< T_{\rm EW}$ where finite temperature effects become important. We also do not consider any possible contribution from the backreaction sector  as this would be impossible to compute model-independently. Consequently, our final result for the relaxion abundance will be a conservative lower bound and the cosmological bounds we derive can possibly be even stronger. For $T\lesssim 20$\,GeV, the dominant production processes are the Primakoff process $q(g)+\phi \to q(g)+ \phi$, involving the $\phi gg$ vertex and the Compton photoproduction process $q+\phi \to q+ \phi$ which involves the $\phi qq$ vertex.  Using the  production rate for the Primakoff process  computed in ref.~\cite{masso}, we get,
\begin{eqnarray}
\Gamma_P= 0.3 \frac{\alpha_s^3 s^2_\theta T^3}{\pi^2 v^2}\,,
\end{eqnarray}
where we have considered only the top loop for computing the $\phi gg$  coupling as the loop contribution of lighter quarks vanishes for temperatures above their masses. For the Compton process  the thermally averaged rate is given by~\cite{turner},
\begin{eqnarray}
\Gamma^f_C\simeq \frac{\alpha_s s^2_\theta T \sum_f m_f^2}{\pi^2 v^2}.
\label{compton}
\end{eqnarray}
Clearly, the dominant contribution is from bottom quarks and the contribution from lighter quarks is negligible. The interference between the Primakoff and Compton processes also  scales as $m_f^2 T$, but is suppressed by another power of  $\alpha_s$ with respect to $\Gamma^f_C$ in \eq{compton} and thus we ignore this contribution. We also  ignore any contribution from the electromagnetic counterpart of the above  processes (that is replacing  gluons by photons in the respective diagrams)  which are expected to be suppressed by powers of $(\alpha_{\rm em}/\alpha_s)$. Thus,  we finally obtain for the total production rate,
\begin{eqnarray}
\Gamma=\Gamma_P+\Gamma_C\,.
\end{eqnarray}

With the knowledge of $\Gamma$ we can now compute the abundance of thermally produced relaxions by solving the Boltzmann equation,
\begin{eqnarray}
Y'=\frac{\Gamma}{x H_t}\left(\frac{0.278}{g_*}-Y\right)\, , \label{boltz}
\end{eqnarray}
where $x=1/T$ and  the Hubble scale $H_t=\frac{\sqrt{4 \pi^3 g_*(T)}}{45}\frac{T^2}{\mpl}$.
Integrating the above, we get
\begin{eqnarray}
Y_{h\phi}&=&Y^{\rm pr}\left[1-\exp\left(-\int^{1/T_f}_{1/T_{0}} {\frac{\Gamma_P}{x H_t}}dx \right)  -\sum_f\left( \int^{1/m_f}_{1/T_{0}} {\frac{\Gamma_C^f}{x H_t}}dx \right)\right]\nonumber\\[3mm]
&\simeq&0.003\left[1-\exp\left(-9\times 10^{11} s_\theta^2\right)\right] \, ,
\label{hp1}
\end{eqnarray}
where $Y^{\rm pr}={0.278}/g^{\rm pr}_*$ and $g^{\rm pr}_*\simeq 86.25$ is the number of relativistic degrees of freedom in energy density in the 1-20\,GeV temperature range. In   the sum over fermion species we include only the $c$ and $b$ quarks as the contribution due to the other quarks is negligible  (see \eq{compton}). We have taken the final temperature $T_f=1$\,GeV for  Primakoff production to justify our use of perturbative QCD  and  $T=m_f$   for the Compton process because below this temperature  the respective fermions become non-relativistic. For   $s_\theta \gtrsim 10^{-6}$ the relaxions have an equilibrium density given by $Y=Y_{eq}=0.003$ whereas for $s_\theta \lesssim 10^{-6}$, the relaxions  have a much smaller density,
\begin{eqnarray}
Y_{h\phi}=2.9 \times 10^{9} s_\theta^2\,.
\label{hp2}
\end{eqnarray}

Once the Universe cools down to a temperature below the quark/hadron transition, i.e.\ $T\lesssim 200$\,MeV, relaxions can be produced via the pion-relaxion conversion process, i.e.\ $N+\pi\to N+ \phi$, $N$ being a nucleon. Using $g_{N\phi}=\frac{m_N s_\theta}{v}$ and $n_N=\left(\frac{m_N T}{2 \pi}\right)^{3/2}e^{-\frac{m_N}{T}}$ we obtain the following parametric estimate for this process,
\begin{eqnarray}
\Gamma_{\pi\phi}\simeq \left(\frac{m_N T}{2 \pi}\right)^{3/2}e^{-\frac{m_N}{T}}\frac{m_N^2 s_\theta^2 T^2}{4 \pi v^2 m_\pi^4}\,.
\label{Nphi}
\end{eqnarray}
One can check that
\begin{eqnarray}
\left.\frac{\Gamma_{\pi\phi}}{H_t}\right|_{T\lesssim 200{\rm~MeV}}\ll\left.\frac{\Gamma_P+\Gamma_C}{H_t}\right|_{T\gtrsim 1{\rm~GeV}}
\end{eqnarray}
and hence we ignore this contribution. Finally, inverse decays may become significant at temperatures just a bit larger than the relaxion mass. The ratio $\Gamma_\phi/H_t$, $\Gamma_\phi$, being the relaxion decay width, is maximal for $T\gtrsim \mphi/5$ as below this temperature, the relaxions become non-relativistic and the rate is Boltzmann-suppressed while above these temperatures $H_t$ increases. We check numerically that
\begin{eqnarray}
\left.\frac{\Gamma_{\phi}}{H_t}\right|_{T=m_\phi/5}\ll\left.\frac{\Gamma_P+\Gamma_C}{H_t}\right|_{T\gtrsim 1{\rm~GeV}}
\end{eqnarray}
and thus the contribution from inverse decays can also be safely ignored.

We now show that the contribution to relaxion abundance from the $q(g)+\phi \to q(g)+ \phi$ processes in \eq{hp1} by far  dominates over the contributions in \eq{mis} and \eq{H2}. First note that we can rewrite \eq{H2} as
\begin{eqnarray}
Y_{H^2}\simeq 2 \times 10^{6} s_\theta^4 \left(\frac{3 {\rm~TeV}}{\Lambda}\right)^{12} \left(\frac{1}{16 \pi^2 r}\right)^4
\end{eqnarray}
using \eq{deriv} and \eq{massmixj}. As we will discuss in detail in the next subsection, cosmological probes are sensitive only if the relaxion decays after 1\,s. As one can see from figure~\ref{mev}, in the region of parameter space which lies in  the untuned area defined in \eq{sthetaH}, if the relaxion decay time is greater than 1\,s (below the purple curve) we must have $\sin \theta<10^{-4}$. In this region $Y_{H^2}$ is clearly always smaller than $Y_{h\phi}$ in \eq{hp1}, even for a cut-off as low as 3\,TeV\@.  As far as the contribution from misalignment, $Y_{m}$, is concerned we have checked numerically that  $Y_{m}\ll Y_{h\phi}$ except in a region of the parameter space where none of the cosmological constraints apply as $Y_{h\phi}<Y_{m}<10^{-20}$ are both extremely small. Thus we conclude that, under our assumptions, the abundance is well approximated by \eq{hp1}.

\subsubsection{Cosmological bounds on late decays}

In this subsection we study the bounds on late decays of the relaxion. The earliest the relaxion has to decay to  have any effect on cosmology is after 1\,s, that is at the neutrino decoupling time, which in the relaxion parameter space  corresponds to  $m_{\phi} < 150$\,MeV  as shown in figure~\ref{kev}. On the other hand for relaxion masses $m_{\phi} < 0.1$\,keV, \eq{sthetaH} implies that $\st \lesssim 10^{-17} $ and thus a lifetime, $\tau_\phi \gtrsim 10^{26}$\,s (see figure~\ref{fig:ctau}) much greater than the age of the Universe ($10^{17}$\,s). This means that for masses $m_{\phi} < 0.1$\,keV  an exponentially small number of relaxions have decayed by the present time and, as we will soon show more rigorously, there are consequently no bounds in this region.
To compute the various constraints from late decays it is important first to know whether the relaxion decays relativistically or non-relativistically at a given point in the parameter space.   If the relaxions are relativistic, their temperature can be computed from their number density,
\begin{eqnarray}
n_\phi&=&Y_\phi s=\frac{\zeta(3)}{\pi^2}T_\phi^3\nonumber\\
\Rightarrow T_\phi&=&\left(\frac{g^S_*}{g_{*}^{\textrm{pr} } } \frac{Y_\phi}{Y^{\rm pr}_{eq}}\right)^{1/3}T_\gamma\,.
\label{tempphi}
\end{eqnarray}
If $T_\phi(\tau_\phi)$,  the relaxion temperature  at the time of its decay, is smaller than $m_{\phi}/5$, it can be safely considered  to have become non-relativistic before decaying. If it becomes non-relativistic,  it can even dominate the energy density of the Universe before decaying (as the energy density of non-relativistic matter decreases more slowly compared to that of relativistic matter). As we will see in this section, such a scenario is highly constrained. In most of the parameter space where various bounds on late decays are relevant, the relaxion decays non-relativistically and thus its energy density before decaying is $\rho_\phi =m_\phi Y_\phi s$. Thus the various  bounds on late decays generally put an upper bound on $m_\phi Y_\phi $ as  a function of the lifetime $\tau_\phi$. Let us now discuss the various constraints on the relaxion decays.

\paragraph{Entropy injection:} If the relaxions decay  after the neutrinos have fully decoupled, i.e.\ for $\tau_\phi\gtrsim 1$ s,  they increase the  entropy of the SM plasma by $\Delta S$,
\begin{eqnarray}
\frac{S_{\rm after}}{S_{\rm before}}=1 +\frac{\Delta S}{S}
\end{eqnarray}
and thus decrease both the baryon-to-photon ratio $\eta_B$ and the effective number of neutrino species, $\neff$. Let us now proceed to compute $\Delta S/S$. For $\tau_\phi\gtrsim 1$ s, relaxions  decay non-relativistically except in a small region of the parameter space with $\st \gtrsim 10^{-4}$  and $\mphi<1$\,MeV  which is outside the region of interest defined in \eq{sthetaH}. In any case for relativistic decays,
\begin{eqnarray}
\frac{\Delta S}{S}\simeq\frac{\rho_\phi}{T_\gamma s}=\frac{3}{4 g^S_*}\left(\frac{T_\phi}{T_\gamma}\right)^4\lesssim 0.3 \%
\end{eqnarray}
and, as we will see, entropy injection smaller than a few percent is unconstrained. To obtain the last inequality above we have used  \eq{tempphi}.  In the rest of the parameter space where the relaxion decays non-relativistically, we must differentiate between the scenario where the relaxion energy density as a fraction of the energy density of  radiation, i.e.,
\begin{eqnarray}
\delta=\frac{\rho_\phi}{\rho_{rad}}=\frac{4}{3}\frac{ g^S_*}{ g_*}\frac{ m_\phi Y_\phi}{ T_\gamma(\tau_\phi)}
\end{eqnarray}
is smaller than unity, $\delta\lesssim1$, from the scenario, $\delta\gtrsim1$,  where the relaxion dominates the energy density.
The entropy injection is given by
\begin{eqnarray}
\frac{\Delta S}{S}=x\cdot\, \left(g^S_*\right)^{1/4}~m_\phi Y_\phi \sqrt{\frac{\tau_\phi}{\mpl}}\,,
\end{eqnarray}
where $x=1.50$~\cite{ishida} for  $\delta\lesssim1$ whereas $x=1.83$~\cite{kolb} for $\delta\gtrsim1$.

Having obtained the expression for $\Delta S/S$, let us proceed to derive the constraints from $\eta_B$ and  $\neff$ measurements.  We first discuss the bound from $\neff$.  Entropy injection anytime after neutrino decoupling and before recombination leads to the reduction in $\neff $, that is:
\begin{eqnarray}
N_{\text{eff}} = 3.046 \left(  \frac{ S_{\text{before}} }{S_{\text{after}}} \right)^{4/3}
\end{eqnarray}
with $\neff^{\textrm{SM}}=3.046$.
Following ref.~\cite{planck} we use the bound $\neff>2.6$ and show in pink  the region excluded by this constraint in figure~\ref{mev} and figure~\ref{kev}.

We now discuss bounds arising from the decrease in the baryon-to-photon ratio, $\eta_B$, caused by relaxion decays.
Since the baryon-to-photon ratio is inversely proportional to  $S$,  $ \eta_B$ is reduced as follows due to entropy injection,
\begin{eqnarray}
\frac{ \eta_{\text{after}} }{\eta_{\text{before}}}= \frac{ S_{\text{before}} }{S_{\text{after}}}.
\end{eqnarray}
A change of $ \eta_B$ between BBN and CMB epoch is not supported by observation since the measured value of $ \eta_B$ during the CMB epoch agrees well with the value after the end of BBN\@. Therefore,   entropy release between these two epochs must  be suppressed. In particular,
CMB  and BBN data constrain $\Delta S/S$ to be smaller than 2$\%$~\cite{Poulin:2015opa}. In figure~\ref{mev} and~\ref{kev} we show the regions of parameter space excluded by this bound in orange.

\paragraph{Big-bang nucleosynthesis:} Big-bang nucleosynthesis (BBN), the formation of light elements in the early Universe, might be altered by
late relaxion decays into SM particles.
The effect depends strongly on the relaxion mass, particularly whether or not it is heavy enough to cause electromagnetic or hadronic cascades.
In our region of interest (i.e.\ for $f>m_h$) relaxions above the pion threshold have a lifetime bigger than 1\,s (see figure~\ref{kev}), so they do not affect cosmology.
Decays of lighter relaxions give rise to electromagnetic showers as long as their mass is bigger than twice the minimum photo-disintegration energy of light nuclei ($\mphi \gtrsim 5$\,MeV)\@.  In the relaxion parameter space (see the beginning of section~\ref{lab}) $m_{\phi} < 150$\,MeV for $\tau_\phi>1$s, so we obtain our bounds from  relaxion decays into electrons.
BBN bounds put constraints on $\rho_\phi/s=m_\phi Y_\phi$ as a function of the lifetime $\tau_\phi$. We consider here the bounds presented in ref.~\cite{Poulin:2015opa} for the decay of a 140\,MeV\,scalar.

The region $f<m_h$ in figure~\ref{kev}, while not relevant for relaxion models, can be interesting in general Higgs portal models. This region can be constrained, for instance, by BBN bounds on decays to pions, hadronic showers etc which can be easily derived using our expression for the abundance in \eq{hp1}.

\paragraph{Distortion of the CMB spectrum:} The energy spectrum of the cosmic microwave background (CMB) allows also to  constrain energy release in the early Universe.
Constraints from CMB distortions become effective for relaxion decays that take place after  $10^6$\,s  as at  earlier times the thermalization process is very efficient.
There are two types of distortions: $ \mu$-distortions and $y$-distortions which dominate at different times.
At  $\tau_{DC}=10^6$\,s ($T_\gamma\sim750$\,eV),  the photon number changing double Compton scattering process ($\gamma+e \rightarrow \gamma+\gamma +e$) freezes out.  As a result,  the photons can no longer be in a Planck distribution (where the number of particles is fixed by the total energy). On the other hand, the Compton process is  active until $\tau_C=10^9$\,s, thus the photons can  still maintain a Bose-Einstein (BE) distribution, but  with a chemical potential $\mu$, whereas the observed Planck spectrum corresponds to an almost vanishing chemical potential. Therefore, $|\mu| $ is constrained by the COBE/FIRAS data which give a bound of
 $|\mu| < 0.9  \times 10^{-4} $ at $95\%$ CL.
 The chemical potential generated by these late decays can be computed to be~\cite{redondo},
\begin{eqnarray}
\mu \simeq  \frac{1}{0.714}  \left( 3  \frac{\rho_{\phi}}{\rho_{\gamma}} - 8 \frac{n_{\phi}}{n_{\gamma}} \right) \left(\exp(-\tau_{DC}/\tau_\phi)-\exp(-\tau_{C}/\tau_\phi)\right)\,.
\end{eqnarray}
In the above equation the factor involving exponentials accounts for the fact that only decays in the time period between $\tau_{DC}$ and $\tau_{C}$ contribute to $\mu$-distortions. If the fractional energy $\delta\ll 1$, one can use $\rho_{\phi}= m_{\phi}  Y_{\phi} s$,  and $\rho_{\gamma}=  \frac{\pi^2}{15} T_\gamma^4$ to find  the constraints whereas the region $\delta \gtrsim 1$  is excluded as it will lead to an ${\cal O}$(1) value for $\mu$ which is excluded. We find that a large  portion of the parameter space is excluded by this constraint as shown in figure~\ref{kev} in green.

If the relaxion decays later than $\tau_C=10^9$\,s ($T \sim 25$\,eV), even the Compton process  freezes out and  this leads to a deviation of the CMB spectrum from a BE distribution. The degree of thermalization that the photons can still achieve can be parametrized by $y$~\cite{redondo},
\begin{eqnarray}
\exp(4 y)-1=\frac{\rho_\phi}{\rho_\gamma} \left(\exp(-\tau_{C}/\tau_\phi)-\exp(-\tau_{\rm RC}/\tau_\phi)\right).
\end{eqnarray}
The region with $\delta \gtrsim 1$ is directly excluded whereas in the region $\delta\ll 1$ we use $\rho_{\phi}= m_{\phi}  Y_{\phi} s$,  and $\rho_{\gamma}=  \frac{\pi^2}{15} T_\gamma^4$ to compute the bound.
In figure~\ref{kev}, we show the region excluded by the bounds from $\mu $ distortions in a darker shade of green than the one denoting $y$ distortions. We also show by dashed lines the projection for the region PIXIE can exclude at 5-sigma level, given by  $|\mu| < 1  \times 10^{-8}$ and $|y| < 5\times 10^{-8}$~\cite{pixie}.

\paragraph{EBL and reionization:} After recombination ($ \tau_{\rm RC} \sim 10^{13}$\,s) the nuclei capture almost all the electrons to form neutral atoms so that the Universe becomes nearly transparent to radiation. The photons injected by relaxion decay can be in principle directly detected, unless their wavelength lies in the ultraviolet range (13.6\,eV-300\,eV)  and they are absorbed in the photoionization process of atoms.  In this ultraviolet mass range bounds from reionization can be set. Photons emitted from very late decays that do not lie in this range,  can be observed today as a distortion of the diffuse extragalactic background light (EBL). The above constraints  can be used to bound the quantity  $\mphi Y_\phi/\tau_\phi$ of  as a function of $\mphi$. Together these bounds cover the wavelength range between 0.1 and 1000 $\mu$m, that is roughly the mass range between 0.1\,eV and 1\,keV\@. We show in figure~\ref{kev} the excluded region using the bounds derived in ref.~\cite{redondo} and~\cite{ringwald}, but appropriately  rescaled to  the different abundance in our case.

\paragraph{Dark matter:} if the relaxion decays after $ \sim 10^{17}$\,s it forms a very small component of the present dark matter density.

\enlargethispage{\baselineskip}

\subsection{Astrophysical probes}

\begin{figure}
 \centerline{  \includegraphics[width=0.65\textwidth]{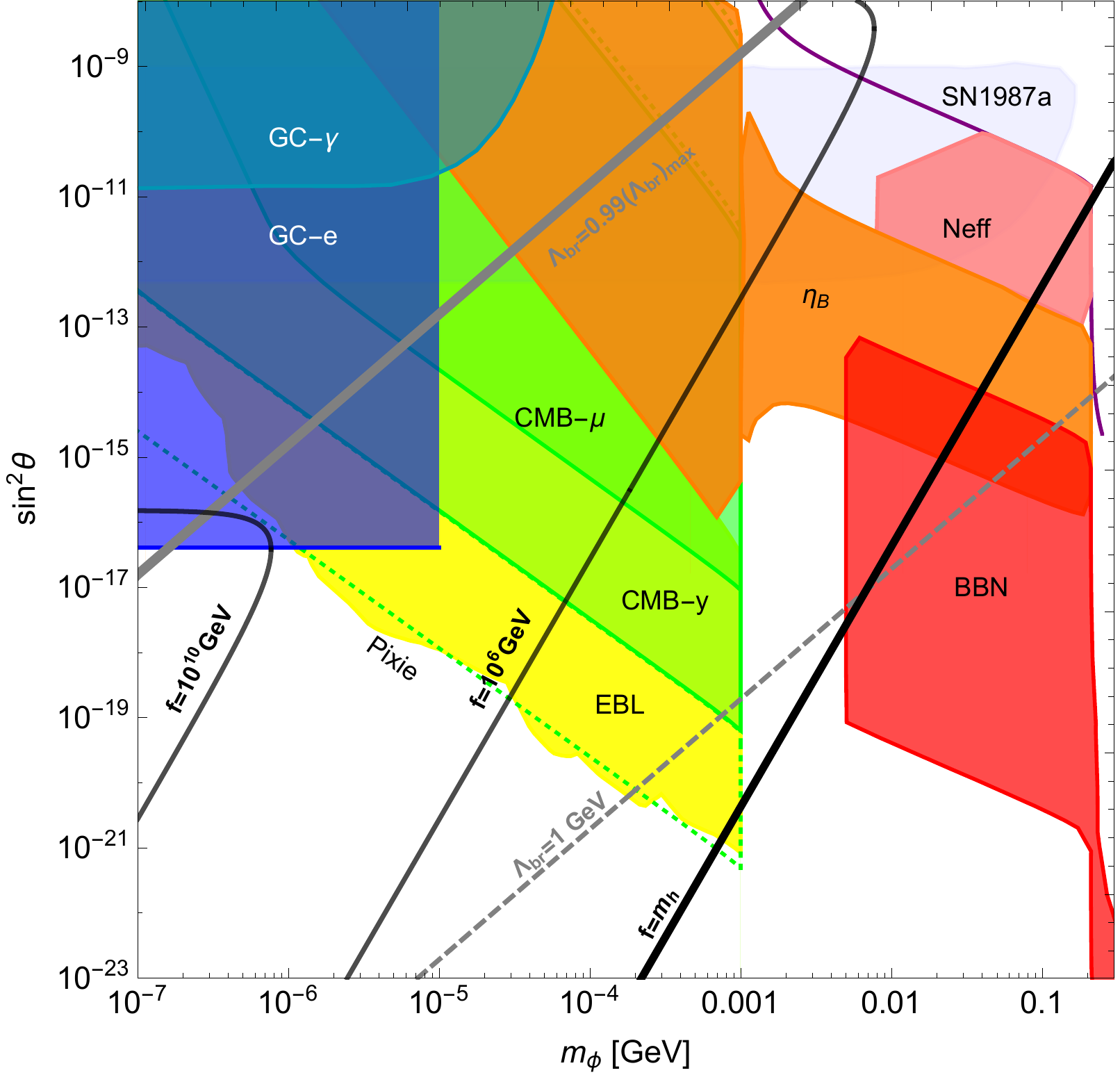}}
\vspace{-3mm}
  \caption{Cosmological and astrophysical bounds on $\sin^2\theta$ and $\mphi$ from $100$\,eV to $0.3\gev$: globular cluster via coupling to electrons (blue) or coupling to photons (turquoise), supernova 1987a (light blue), extragalactic background light (EBL, yellow), CMB $y$-distortion (light green) and $\mu$-distortion (green), entropy injection $\Delta S/S$ bounded by the baryon-to-photon ratio $\eta_B$ (orange) and by $\neff$ (pink), BBN (red). The green dotted lines represent the  projection for the sensitivity of  of PIXIE to CMB distortions.The light gray band indicates the possible range of $\sin^2\theta$ for $j=1$, i.e.\ the QCD case.
  The gray lines (from top to bottom) are contours of constant $\Lambda_{\rm br}=0.99\lbrmax$ (thick, solid), and $1\gev$ (dashed). Here $\lbrmax$  is the upper bound on $\lbr$ arising from the requirement of a non-tachyonic $\phi$ in \eq{tach} for $ \sin (\phi_0/f)=1/\sqrt{2}$. The black lines (from left to right) are contours of constant $f=10^{10}\gev, 10^6\gev$ (thin) and $f=m_h$ (thick).}
  \label{kev}
 \end{figure}

\paragraph{SN1987a supernova:} In the core of a supernova, a relaxion can be produced  via its couplings to nucleons and thereby contribute to its energy loss.
The relevant process is bremsstrahlung $ N  + N \rightarrow N + N + \phi$.
Requiring that the energy loss into the new scalar must be smaller than the measured energy loss  into neutrinos leads to bounds
 on the Higgs-relaxion mixing as long as the relaxion is lighter than 20\,MeV\@. In figure~\ref{kev} and~\ref{mev}   we show (in light blue) the bounds derived in~\cite{Krnjaic:2015mbs}, using the results of ref.~\cite{Ishizuka:1989ts}.  This computation is exponentially sensitive to some uncertainties (see ref.~\cite{Krnjaic:2015mbs}) and thus should be interpreted only as an order of magnitude estimate.  At a more conceptual level, even the idea of energy loss via neutrinos has been questioned in the literature~\cite{Blum:2016afe}. New laboratory constraints  that are able to explore this region are therefore required.

\paragraph{Globular-cluster star bounds:} Relaxions can be produced in globular-cluster (GC) stars via processes involving the relaxion electron coupling, $g_{\phi e}$, such as the  Compton and bremsstrahlung processes. Requiring that the total cooling rate is not  faster than expected~\cite{Agashe:2014kda,Raffelt:2012sp} gives us the bound
\begin{eqnarray}
g_{\phi e} < 1.3  \times 10^{-14}\Rightarrow \st < 4 \times 10^{-17}
\end{eqnarray}
for $ m_{\phi}\lesssim 10$\,keV\@.
Limits can be also set on the relaxion-photon coupling considering Primakoff  photon-relaxion conversion~\cite{Raffelt:2012sp, redondo},
\begin{eqnarray}
g_{\phi \gamma} < 0.6  \times 10^{-10} \;  \text{GeV}^{-1}\Rightarrow \st < 1\times 10^{-11}
\label{gcp}
\end{eqnarray}
for $ m_{\phi}\lesssim 30$\,keV\@. In figure~\ref{kev}  the GC  limit on $g_{\phi e}$ is presented in blue and the one on $g_{\phi \gamma}$  in turquoise.

\paragraph{CAST experiment:} The CERN Axion Solar Telescope (CAST) looks via X-rays for axion-like particles coming from the sun.\enlargethispage{\baselineskip}
 The present limit on the photon-ALP coupling is~\cite{Graham:2015ouw,Dafni:2016xsz}:
 \begin{eqnarray}
g_{\phi \gamma} < 0.8   \times 10^{-10} \;  \text{GeV}^{-1} \Rightarrow \st<2 \times 10^{-11}
\label{cast}
\end{eqnarray}
for  $ m_{\phi}< 0.02 $\,eV\@.
The limit is slightly weaker than the GC limit and well outside the region  of interest in \eq{sthetaH}, hence we omit it in figure~\ref{kev}. In contrast, IAXO~\cite{Ribas:2015hwf}, the new generation experiment, will be able to improve the limit.
However, despite the future progress in this technology this class of experiments is not likely to be relevant for our scenario since it probes a region of the parameter space where fifth force experiments provide very strong bounds.

\section{Implications for the relaxion theory space}\label{sect:summaryplots}

\begin{figure}
\centerline{ \includegraphics[width=0.8\textwidth]{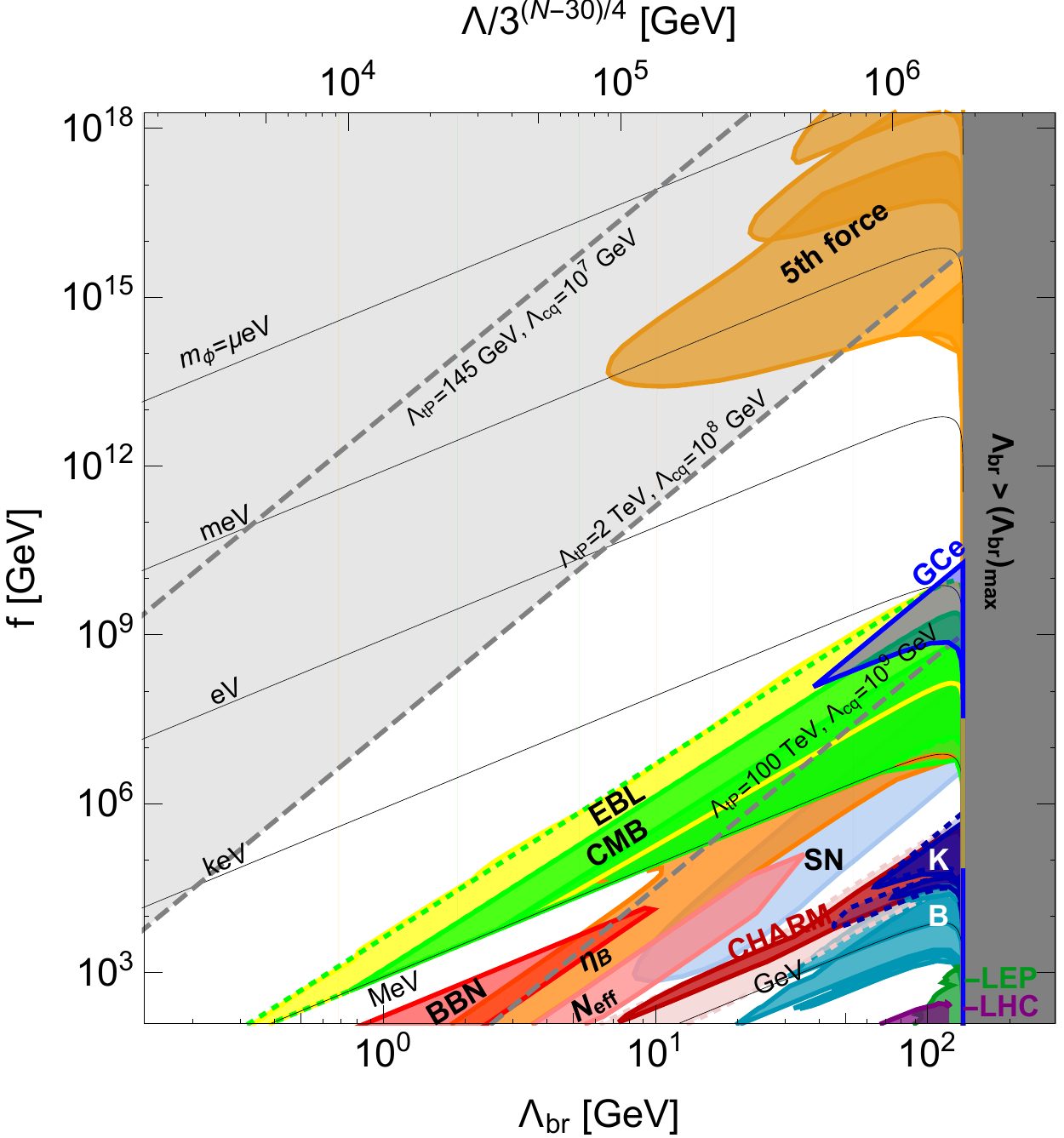}}
\vspace{-2mm}
 \caption{Summary of laboratory, cosmological, astrophysical and collider bounds on $\lbr$ and $f$. The upper horizontal axis bounds the cut-off $\Lambda$ for $N=30$ via Eq.~(\ref{en3}). For other $\Lambda$, the required $N$ is obtained via $N = 4 \log_3\left(\frac{\Lambda}{R}\right) + 30$ where $R=\Lambda/3^{(N-30)/4}$ is the value read off the upper axis.
 Laboratory: fifth force experiments (light orange). Cosmology and astrophysics: EBL (yellow), CMB (green),
 globular cluster via coupling to electrons (blue, transparent), BBN (red),
 entropy injection constrained by $\eta_B$ (orange) and by $\neff$ (pink), supernova 1987a (light blue). The green dotted lines represent the  projection for the sensitivity of  of PIXIE to CMB distortions.
 Beam dump experiments: CHARM (dark red) and projections for SHiP (dark red, dotted).
 For the beam dump projections at NA62 and SeaQuest, see figure~\ref{mev}.
 Flavour: rare $K$-meson decays at E949/787, NA48/2, KTEV (dark blue) and projection for NA62 (dark blue, dotted), rare $B$-meson decays at Belle and LHCb (turquoise).
 Higgs production and decay at colliders: LEP (green), LHC (purple).
 The vertical gray band indicates exclusion due to $\lbr>\lbrmax$ (here $\lbrmax$ is the upper bound on $\lbr$ arising from the requirement of a non-tachyonic $\phi$ in \eq{tach} for $\sin (\phi_0/f)=1/\sqrt{2}$).
 The dashed, black lines show (from top to bottom) contours of $\Lambda=\left\lbrace 145\gev, 2\tev, 100\tev \right\rbrace$ for $\Delta\phi=\mpl$ from the transplanckian (``tp") condition in Eq.~(\ref{trp}). The same contours are obtained for $\Lambda\simeq 10^7\gev, 10^8\gev, 10^9\gev$ from the cosmological classical-vs.-quantum (``cq") condition in Eq.~(\ref{cosmo}).
 The thin, black lines indicate $\mphi$ from $10^{-15}\gev$ (uppermost) to $1\gev$ (lowest) with a spacing factor of $10^3$.
 }
 \label{fig:Mtif}
 \end{figure}

In this section, we collect all bounds from laboratory experiments, colliders, astrophysics and cosmology that were shown in figures~\ref{ev},~\ref{mev},~\ref{gev} and~\ref{kev} for different mass regions and translate them, using \eq{inv}, into the underlying theory parameters $\lbr$ and $f$ in figure~\ref{fig:Mtif}.\footnote{The color coding for the experimental bounds is the same as in the previous figures.}
As a connection between both parametrisations, $\lbr$ and $f$ were shown as a grid of contours in the previous plots, whereas in the $(\lbr, f)$ plane of figure~\ref{fig:Mtif} we show contours $\mphi$.
While the values we provide are for the $j=2$ case, as mentioned below \eq{inv} one can obtain the values for the $j=1$ by the simple translation $\lbr\to \sqrt{2} \lbr, f\to 2 f$.

We show how these bounds push the cut-off to smaller values by the upper horizontal axis, where we translate the $\lbr$ scale in the lower axis  to cut-off values using \eq{en2} for $n=3^N=3^{30}$. As indicated in the figure these values can be easily rescaled for other values of $n$ or $ N$.

The overview presented in figure~\ref{fig:Mtif} shows that large areas in the $\lbr-f$ plane are already well covered by existing experimental and observational probes, for instance  the high-$f$ region up to $\mpl$ is probed by the fifth force experiments, on the other hand the cosmological, astrophysical, beam dump and collider observables constrain lower values of $f$. We see that in the above $f$ ranges, the region with electroweak scale $\lbr$ is practically completely ruled out apart from small gaps that still remain. We also show in figure~\ref{fig:Mtif} how some of these gaps in parameter space might be covered soon by future experiments such as SHiP, NA62 and PIXIE.  However, the region between $f\sim 10^{10}\gev$ and $10^{14}\gev$  which corresponds to  relaxion masses  between $0.1$\,eV and 1\,keV, is currently barely  constrained by data.
\enlargethispage{\baselineskip}

For any $f$ (or $m_\phi$) value, all the constraints can be evaded for sufficiently small $\lbr$  values (there are no bounds for $\lbr\lesssim 0.3\gev$). Small $\lbr$  values are however theoretically disfavoured for several reasons. First  of all, as we see from the $\Lambda_{cq}$  contours in figure~\ref{fig:Mtif},
the constraints derived here push the relaxion to a region with somewhat lower values for the upper bound on the cut-off derived from cosmological considerations during inflation. If one takes seriously the requirement that the relaxion should not have transplanckian excursions, our bounds  have a much stronger  impact. This is because, as we see from figure~\ref{fig:Mtif}, our bounds already cover a large part  of the parameter space outside the shaded region where the relaxion travels transplanckian distances for any cut-off larger than 2\,TeV\@. Coming  to the issue of the very large global charges that arises due to the compact nature of the relaxion, we see from the upper horizontal axis that even in CKY/clockwork models the number of sites required can become uncomfortably  large for very small backreaction scales. For $N\lesssim 30$ (see section~\ref{sectheory})  our bounds can significantly constrain the cut-off.  For  instance for $f= 1000$\,TeV we find $\Lambda\lesssim 100$\,TeV\@. As shown in section~\ref{sectheory}  the simplest clockwork  models start getting tuned for $N\gtrsim 30$. As far as the proposal to solve the little hierarchy problem using modest $n$ values is concerned~\cite{familon}, we see that such a proposal would be completely ruled out outside the $f\sim 10^{10}\gev$- $10^{14}\gev$  ($\mphi\sim 0.1$\,eV - 1\,keV) region, as contrary to the philosophy of this approach, too large values of $n>(v/\lbr)^4$ would be required.

One should be keep  in mind while interpreting  these bounds within the clockwork framework that in these models one must have $f\gtrsim  \Lambda$ from \eq{claudiabnd}. Thus even from this point of view  the unconstrained $f\sim 10^{10}\gev$-$10^{14}\gev$  ($\mphi\sim 0.1$\,eV-1\,keV) window is an interesting region as here  the cut-off can be high in these models.

Finally let us discuss what impact the pseudoscalar couplings of the relaxion might have on the overall bounds.
As explained in appendix~\ref{ps},  in the electroweak preserving~\cite{kaplan,familon} models discussed in section~\ref{secbr},  the relaxion does not have pseudoscalar couplings larger than the Higgs-portal ones, hence our experimental bounds would be  qualitatively unchanged.
Let us briefly comment on the possible change in our bounds if the pseudoscalar coupling to photons  is larger than the one induced by Higgs mixing. As already mentioned, among the models discussed in section~\ref{secbr} this holds only for the pseudoscalar diphoton coupling in the non-QCD $ j=1$ model where the relaxion has a pseudoscalar coupling to photons suppressed only by $1/ f$  and not by the backreaction scale (see \eq{gphit1}).
  In this case the astrophysical and cosmological bounds discussed in section~\ref{cosmosec} will be affected. An analysis of how the cosmological bounds change in the presence of a large $\tilde{g}_{\phi \gamma}$ coupling is beyond the scope of this work.
  The enhanced coupling to photons will lead to a stronger bound from globular clusters, that is $ f \gtrsim 10^7$\,GeV, Eq.~(\ref{gcp}).
  However, this is valid only provided that the relaxion mass is lighter than 30\,keV, so we immediately see from figure~\ref{fig:Mtif} that this is relevant only for  $\Lambda_{\rm br}\ll v$.
   Furthermore, the CAST experiment can put a bound on $f$ of similar order on the coupling to photon Eq.~(\ref{cast}) in the sub-eV region.  For  large Higgs-relaxion mixing fifth force experiments are sensitive, hence the CAST bound is irrelevant. However for $\Lambda_{\rm br}\ll v$, when the sensitivity to fifth force experiments ceases, the CAST bound on the pseudo-scalar coupling can be important for sub-eV relaxions.

\section{Testing for the CP violating nature of the relaxion}
\label{cpv}

In this section we investigate
the feasibility of detecting a signal of spontaneous CP-violation together with a Higgs mixing signal. This would represent a smoking gun for our scenario since
what we discussed so far about relaxion phenomenology applies to any scalar mixed with the Higgs.
However, as already discussed in appendix~\ref{ps}, the strength of the relaxion pseudo-scalar couplings depend on the details of the  back-reaction sector. Couplings to fermions are typically very suppressed (compared to the one from Higgs-relaxion mixing), while the coupling to photon  $ \tilde{g}_{\phi \gamma} $ is in many cases only as large as the scalar one, that is  $ \tilde{g}_{\phi \gamma} \sim 10^{-5} \sin{\theta}$.
In the electroweak breaking non-QCD model discussed in section~\ref{secbr}, instead, the coupling to photons is in principle larger since it is not suppressed by the backreaction scale.
 Despite the model dependence, it is still an interesting question whether a CP-violating signal could be detected at the precision frontier.
Let us investigate the relaxion contribution to the electric dipole moments (EDM). In our scenario the leading contribution to the electric dipole moment  is generated through its couplings to fermions via Higgs mixing and with the pseudoscalar coupling to photons,  $\tilde{g}_{\phi \gamma}$.We will focus on the electron EDM, following~\cite{Marciano:2016yhf}, but similar results hold for the neutron EDM\@.

The first step is to understand in which relaxion mass range this probe can be effective. To this end let us estimate the strength  of $ \tilde{g}_{\phi \gamma}  \times g_{\phi e} $ since the relaxion one-loop contribution to the electron EDM will be proportional to it.
The current  upper bound on the electron EDM is  $  d_e/e \sim 8 \times 10^{-29}  \text{cm}$~\cite{Baron:2013eja}, which corresponds to $\tilde{g}_{\phi \gamma} g_e \sim 5 \times  10^{-14} \,\mbox{GeV}^{-1}$~\cite{Marciano:2016yhf},  and  improvements of one order of magnitude are expected in the coming years~\cite{Hewett:2012ns}.
 Let us then see how this compares to relaxion models.
 For the non-QCD electroweak breaking model we get:
 \begin{eqnarray}
 \tilde{g}_{\phi \gamma} g_{\phi e}  \lesssim  \frac{ \alpha}{ 4\pi} \frac{m_e}{v} \frac{ \sin{\theta}}{f}\sim
 3\times10^{-16}  \bigg(\frac{m_{\phi}}{1 \text{GeV}} \bigg)^2 \; \text{GeV}^{-1} \, ,
 \end{eqnarray}
where we used \eq{eq:gphipsi} and \eq{gphit1}.

The electroweak preserving models~\cite{kaplan,familon} have an additional  $\Lambda^4_{\rm br}/v^4$ suppression from the backreaction scale due to the suppression in $\tilde{g}_{\phi \gamma}$ in \eq{gphit21} as compared to \eq{gphit1}.
We see that in both cases, a relaxion  with $\mphi\simeq 1 \gev$ yields a contribution to the $d_e$ that is below the current (and near future) sensitivity.
The parameter space constrained is therefore in the few GeV region.

\section{Conclusions}\label{conc}

We study various phenomenological aspects of relaxion models. We focus on models where the rolling of the relaxion field stops due to the presence of a Higgs-relaxion backreaction term.
We show that the  relaxion  generically stops its rolling at a point that  breaks the CP symmetry, leading to relaxion-Higgs mixing.
  We investigate then the implications of this mixing, and analyse current and near future probes involving laboratory, cosmological and astrophysical measurements in terms of reach and sensitivity. In most parts of the parameter space, these observational constraints put the most stringent bound on the backreaction scale, $\Lambda_{\rm br}\,$. On the theoretical front, we show that simple multiaxion (clockwork) UV completions suffer from a  fine tuning problem, which increases with the number of sites.

   Let us describe in more detail our main results on the observational probes of relaxion-Higgs mixing.  The constraints/discovery prospects derived by us are summarised in figures~\ref{ev}--\ref{kev}. In the sub-eV mass range the relaxion lifetime is much larger than the age of the Universe and thus cosmological or direct laboratory probes  are not effective. Fifth force experiments, however, are sensitive in large regions of the parameter space in the sub-eV region because of  the  low mass of the relaxion and the CP-even nature of its couplings to SM particles via Higgs mixing (see figure~\ref{ev}). The eV-MeV region is practically unconstrained by laboratory probes, but a subset of this region (keV-MeV) can be constrained by astrophysical and cosmological probes as shown in figure~\ref{kev}. The cosmological probes are relevant here because this is the region of parameter space where the relaxion lifetime is between 1\,s and $10^{26}$\,s and thus is tested by a variety of cosmological probes, such as  entropy injection constraints from $\neff$ and $\eta_B$ measurements, BBN observables, CMB  spectral distortions and EBL distortions. Turning to the MeV-GeV region we find that in some parts of this mass range, the relaxion lifetime is just right for beam dump experiments
   (${\cal O}(100$\,m) in the lab frame) such as the CHARM experiment and experiments probing invisible rare meson decays. We also find that future data from beam dump experiments like SeaQuest and especially SHiP and the currently running ultra-rare kaon decay experiment NA62 can probe new and interesting regions of the relaxion parameter space.   In other parts of this MeV-GeV mass region visible rare meson decays also put significant bounds. Finally, for relaxion masses above 5\,GeV the constraints arise from LEP bounds on the Higgs-strahlung process and LHC Higgs coupling bounds on the new channel, $h \to \phi \phi$, as shown in figure~\ref{gev}.
     In figure~\ref{fig:Mtif}, we translate these bounds to the relaxion theory space and discuss the theoretical implications.
We finally comment that, while the relaxion-Higgs mixing requires CP violation, most of the probes discussed above do not form a strong test of the CP nature of the relaxion. The pseudoscalar couplings of the relaxion tend to be more model-dependent. For instance, in the familon model that was constructed in~\cite{familon} the relaxion does not couple to $F_{\mu\nu}\tilde F^{\mu\nu}$ (with $F$ being the QED field strength) at one loop but only to the orthogonal combination of the electroweak field strengths. We find that,  in existing models, probes of CP violation are sensitive only for GeV scale relaxion masses.

\acknowledgments

We thank Kfir Blum, Josef Pradler,  Gordan Krnjaic and Aviv Shalit for helpful discussions and suggestions. We thank Babette D\"obrich and Gaia Lanfranchi for useful and detailed discussions about NA62. TF would like to thank the Weizmann theory group for hospitality during the initial stages of the project. TF was supported by the Basic Science Research Program through the National Research Foundation of
Korea (NRF) funded by the ministry of Education, Science and Technology (No. 2013R1A1A1062597) and by by the Korea- ERC researcher visiting program through NRF (No. 2015K2A7A1036922) and by IBS under the project code, IBS-R018-D1. The work of GP is supported by grants from the BSF, ERC, ISF, Minerva, and the Weizmann-U.K. Making Connections Programme.

\paragraph{Note added:} as this paper was being completed, ref.~\cite{Choi:2016luu} appeared which also discusses relaxion phenomenology.

\appendix

\section{Relaxion mass and mixing for the case of \texorpdfstring{$\boldsymbol{j=1}$}{j=1}}\label{app1}

In this appendix we present the mass matrix for the $j=1$ case. The potential in this case~is,
\begin{eqnarray} \label{heavy1}
V = \left[- \Lambda^2 +g  \Lambda \phi+\dots\right] \hat{h}^2 -\tilde{M}^{3} \hat{h} \cos \left( \frac{\phi}{f} \right)+\lambda \hat{h}^4
  + rg \Lambda^3  \phi+ rg^2 \Lambda^2  \phi^2 \, . \label{eq:V11}
  \end{eqnarray}
Expanding around their vacuum expectation values (VEVs),
\begin{eqnarray}
\phi=\phi_0+ {\phi'}\,,~~~~~~\hat h=\frac{v_H+h'}{\sqrt{2}}\,,
\end{eqnarray}
and imposing the minimisation conditions, we get
\begin{eqnarray} \label{minim11}
 \lambda v_H^2 -\Lambda^2 + g \Lambda \phi_0- \frac{\tilde{M}^3}{\sqrt{2}v_H} \cos\left(\frac{\phi_0}{f}\right)&=&0\,, \\
r g \Lambda^3+2 r g^2 \Lambda^2 \phi_0 +\frac{g \Lambda v_H^2}{2}+  \frac{\tilde{M}^3 v_H}{\sqrt 2f}   \sin \left(\frac{\phi_0}{f}\right)&=&0\,, \label{minim21}
 \end{eqnarray}
where $\phi_0\sim \Lambda/g$ and $\phi_0/f$ is expected to be an ${\cal O}(1)$ phase. Now we calculate the $\phi'-h'$ mass matrix,
\begin{eqnarray}
M^2_{h'h'}&\equiv&  \frac{\partial^2 V}{\partial {h'}\partial {h'}} =3 \lambda v_H^2 -\Lambda^2 + g \Lambda \phi=2 \lambda v_H^2+  \frac{\tilde{M}^3}{\sqrt{2}v_H} \cos\left(\frac{\phi_0}{f}\right)\,, \\
M^2_{h'{\phi'}} &\equiv& \frac{\partial^2 V}{\partial h \partial {\phi'}} = g\Lambda v_H+ \frac{\tilde{M}^3}{\sqrt{2}f} \sin\left(\frac{\phi_0}{f}\right)\simeq  \frac{\tilde{M}^3}{\sqrt{2}f}\sin\left(\frac{\phi_0}{f}\right)\label{hph1}\,,\\
M^2_{{\phi'}{\phi'}}&\equiv& \frac{\partial^2 V}{\partial {\phi'}\partial {\phi'}} =  \frac{\tilde{M}^3v_H}{\sqrt{2}f^2} \cos\left(\frac{\phi_0}{f}\right)+2r g^2 \Lambda^2 \simeq  \frac{\tilde{M}^3v_H}{\sqrt{2}f^2} \cos\left(\frac{\phi_0}{f}\right)\,,\label{php1h}
\end{eqnarray}
where  we have used  \eq{minim11}, \eq{minim21} and   $\Lambda^2\gg r v_H^2$ to obtain the approximations above. For any given $\mti$ (or $\lbr$) and $f\,$ the exact relaxion mass and mixing can be determined by diagonalising the above mass matrix after setting $\lambda$ by requiring the heavier eigenvalue to be the physical Higgs mass, $m_h=125\gev$. Note that we always have,
\begin{eqnarray}\label{mincon}
m_h^2 >M^2_{h'h'}=2 \lambda v_H^2+  \frac{\tilde{M}^3}{\sqrt{2}v_H} \cos\left(\frac{\phi_0}{f}\right),
\end{eqnarray}
a fact we use in section~\ref{secbr}\,.

\section{Relaxion mass and mixing for the case of \texorpdfstring{$\boldsymbol{j=2}$}{j=2}}
\label{app2}
In this appendix we derive the relevant relation for $j=2$ models. To obtain this we expand the potential $V(h, \phi)$,
\begin{eqnarray} 
V= \left[- \Lambda^2 +g  \Lambda \phi+\dots\right] \hat{h}^2 -\tilde{M}^{2} \hat{h}^2 \cos \left( \frac{\phi}{f} \right)+\lambda \hat{h}^4
  + rg \Lambda^3  \phi+r g^2 \Lambda^2 \phi^2 \dots \label{eq:V1}\,.
  \end{eqnarray}
 around the minimum $(v_H, \phi_0)$. In these models $v_H=v=246$\,GeV\@. The minimisation conditions yields,
\begin{eqnarray} \label{minim1}
 \lambda v^2 -\Lambda^2 + g \Lambda \phi+g^2\phi^2- \tilde{M}^2 \cos \left(\frac{\phi_0}{f}\right)&=&0 \,,\\
r g \Lambda^3 +\frac{g \Lambda v^2}{2}+  \frac{\tilde{M}^2 v^2}{2f}   \sin \left(\frac{\phi_0}{f}\right)&=&0\,, \label{minim2}
 \end{eqnarray}
where $\phi_0 \sim \Lambda/g$ as in section~\ref{intro}, and the trigonometric functions have ${\cal O}(1)$ values. The $\hat{\phi}-h$ mass matrix results in
\begin{eqnarray}
M^2_{h'h'}&\equiv&  \frac{\partial^2 V}{\partial {h'}\partial {h'}} =3 \lambda v^2 -\Lambda^2 + g \Lambda \phi- \tilde{M}^2 \cos \left(\frac{\phi_0}{f}\right)= 2  \lambda v^2\,, \nonumber\\
M^2_{h'{\phi'}} &\equiv& \frac{\partial^2 V}{\partial h' \partial {\hat{\phi}}} = g\Lambda v+\frac{\tilde{M}^2v}{f} \sin\left(\frac{\phi_0}{f}\right)\simeq  \frac{\tilde{M}^2v}{f} \sin\left(\frac{\phi_0}{f}\right)\,, \nonumber \\
M^2_{\phi'\phi'}&\equiv& \frac{\partial^2 V}{\partial {\phi'} \partial {\phi'}} = \frac{v^2}{2} \frac{\tilde{M}^2}{f^2} \cos\left(\frac{\phi_0}{f}\right)+2 r g^2 \Lambda^2 \simeq  \frac{v^2}{2} \frac{\tilde{M}^2}{f^2} \cos\left(\frac{\phi_0}{f}\right)\label{phph}\,.
\end{eqnarray}
where  we have used  \eq{minim1}, \eq{minim2} and   $\Lambda^2\gg r v^2$ to obtain the approximations above. For any given $\mti$ (or $\lbr$) and $f$, the exact relaxion mass and mixing can be determined by diagonalising the above mass matrix after setting $\lambda$ by requiring the heavier eigenvalue to be the physical Higgs mass, $m_h=125$\,GeV\@.
\section{Pseudoscalar couplings of the relaxion}
\label{ps}

In this appendix we discuss the pseudoscalar couplings of the  relaxion that can arise from the backreaction sector. As already mentioned in section~\ref{secmix}, these couplings are model-dependent so our discussion here would be limited to  the specific models   in section~\ref{secbr}, namely the $j=1$ non-QCD backreaction model and the $j=2$ axion-like and familon models. As far as  low energy probes are concerned, the important couplings are the ones to light fermions, photons and gluons. In all the above models, the exotic fermions\footnote{In the non-perturbative $j=1,2$ models it is the analog of the pion and /or  $\eta'$ that get the loop induced couplings (to both fermions and photons) and the relaxion obtains its coupling via mixing with these states.} can induce a pseudoscalar coupling of the relaxion to light fermions, $\tilde{g}_{\phi f}$,  that is proportional to the light fermion mass as well as the shift symmetry breaking spurion ($\sim \lbr^4$) that generates the relaxion mass; thus it has the same suppressions as the Higgs portal coupling, ${g}_{\phi f}$ in section~\ref{secmix}. Furthermore, as the above models involve sequestered sectors, one can check by inspection that  these couplings are generated at least one loop order higher than the corresponding Higgs-portal coupling so that
\beq
\tilde{g}_{\phi f}\sim \frac{{g}_{\phi f}}{16 \pi^2}.
\eeq

As far as the coupling to photons is concerned we need to distinguish between the $j=1$ and $j=2$ models. By inspection we see that in both the $j=2$ models  the $\frac{\tilde{g}_{\phi \gamma}}{4}\phi F \tilde{F}$ coupling can be possibly induced  but only with the same shift symmetry breaking suppression ($\sim \lbr^4/v^4$) and at the same  order in perturbation theory as the Higgs portal coupling (in the non-perutbative model the coupling can arise via mixing with the analog of the  $\eta'$ and in the perturbative familon model via a 2-loop level diagram),
\beq
\tilde{g}_{\phi \gamma} \sim \frac{\Lambda_{\rm br}^4}{v^4}\frac{\alpha_{em}}{4 \pi f} \sim {g}_{\phi \gamma}.
\label{gphit21}
\eeq

 In the non-QCD $j=1$ model, however, it is possible to have $\tilde{g}_{\phi \gamma}\gg {g}_{\phi \gamma}$  because this backreaction sector is just a scaled-up version of QCD\@. Thus, as is the case for QCD axions,  the relaxion will get an anomaly-induced coupling of the same order via mixing with the $\eta'$ and pion analogs of the new strong sector. This generates
  \begin{eqnarray}
 \tilde{g}_{\phi \gamma} \sim \frac{\alpha_{em}}{4 \pi f}
 \label{gphit1}\,,
 \end{eqnarray}
 which can be larger than the Higgs portal coupling, ${g}_{\phi \gamma}$,  for  values of $\Lambda_{\rm br}\ll v$.

It is important to mention that while the pseudoscalar coupling of the relaxion to photons is smaller than the Higgs-portal one in the existing $j=2$ models, an anomaly induced  coupling of the size in \eq{gphit1} would exist in simple variants where the relaxion couples directly to the the electroweak doublet fermions.

 One can  proceed along the same lines to show that the pseudoscalar coupling of the relaxion to gluons is at least one loop suppressed with respect to the Higgs portal induced coupling to gluons because of  the sequestering. We see, therefore, that apart from the $ \tilde{g}_{\phi \gamma}$ coupling in the $j=1$ model, the pseudoscalar couplings of the relaxion are either suppressed or of the same order as the Higgs portal coupling in the models in section~\ref{secbr}. Our results would thus be qualitatively unchanged by the presence of these couplings apart from the one exception above, on which we comment  in the text.

\section{The \texorpdfstring{\boldmath $h\phi\phi$ coupling in $j=2$}{h phi phi coupling in j=2} models}\label{apphpp}
In this appendix we present the expression for the $h\phi\phi$ coupling in $j=2$ models. To obtain this we expand the potential $V(h, \phi)$ in \eq{eq:V} around the minimum $(v, \phi_0)$ to obtain all cubic terms and then substituting the gauge eigenstates in terms of the mass eigenstates
\begin{eqnarray}
\phi'=-\sint\, h + \cost\, \hat \phi\,,\nonumber\\
h'=\cost\, h +\sint\, \hat \phi\,.
\end{eqnarray}In order to reduce the complexity the full expression while accounting for the leading mixing effects, we take $\sin ({\phi_0}/{f})=\cos ({\phi_0}/{f})=1/\sqrt{2}$  to finally obtain
\begin{align}
 g_{h\phi\phi} &\simeq \frac{\tilde M^2}{\sqrt{2}f}\left(
 -\frac{ v^2 s_{\theta} c^2_{\theta}}{4 f^2}
 +\frac{v c^3_{\theta}}{2 f}
 -\frac{v s^2_{\theta} c_{\theta}}{f}
 +\frac{ s^3_{\theta}}{2 }
 - s_{\theta} c^2_{\theta}\right)
 +3 \lambda  v s^2_{\theta} c_{\theta}
 \label{eq:gmix}\,,
\end{align}
where all the terms proportional to powers of $g$ such as the leading contribution $\Delta g_{h\phi\phi} \sim g \Lambda s_\theta c_\theta^2$ can be shown to be sub-dominant compared to the terms in \eq{eq:gmix}, using \eq{minim2} and assuming $\Lambda^2 \gg v^2$.  We use this expression to derive bounds on the decay of $h\to \phi\phi$, $\br(h\rightarrow \phi\phi)$ in section~\ref{lep}. In a similar manner we can derive the $h \phi \phi$ coupling for the $j=1$ case. While we do not perform the full computation here, we note that the leading term in that case would be $g_{h\phi\phi} \sim \tilde{M}^3 c_\theta^3/f^2$.

\section{Expressions for relaxion partial widths and lifetime}\label{appw}

In this appendix we provide the expressions for the relaxion partial  widths for different channels. The dilepton ($l\bar l$) and diphoton ($\gamma\gamma$) partial widths are given by
 \bea
 \Gamma( \phi \rightarrow l \bar l) &=&\st \frac{m_l^2}{v^2} \frac{m_{\phi}}{8 \pi}  \left(1 -  \frac{4 m_l^2}{m_{\phi}^2}\right)^{3/2}\,,\nonumber\\
  \Gamma( \phi \rightarrow \gamma \gamma) &=&\st \frac{g^2_{\phi \gamma}\mphi^3}{64 \pi} \, .
 \eea
 As far as colored states are concerned we use the perturbative description above  $\mphi=1\gev$. The partial width to quarks ($q\bar q$) and gluons ($gg$) is given by
  \bea
 \Gamma( \phi \rightarrow q \bar q) &=&\st \frac{3 m_q^2}{v^2} \frac{m_{\phi}}{8 \pi}  \left(1 -  \frac{4 m_q^2}{m_{\phi}^2}\right)^{3/2} \, ,\nonumber\\
  \Gamma( \phi \rightarrow gg) &=& \frac{g^2_{\phi g}\mphi^3}{8 \pi}\, , \nonumber\\
  \Gamma( \phi \rightarrow \gamma \gamma) &=& \frac{g^2_{\phi \gamma}\mphi^3}{64 \pi}.
 \eea
For $\mphi<1\gev$ the only hadronic state we consider is the decay to pions. Different estimates of the partial  width to pions vary over nearly two orders of magnitude~\cite{Clarke:2013aya}. Here we use the leading order calculation of ref.~\cite{voloshin} which gives
 \bea
 \Gamma( \phi \rightarrow \pi \pi) &=&\st\frac{3}{32 \pi v^2 \mphi}  \left(1 -  \frac{4 m_\pi^2}{m_{\phi}^2}\right)^{1/2}\left(\frac{2 \mphi^2 +11 m_\pi^2}{9}\right)^2.
  \eea
  For $\mphi>1\gev$ one should use the partial width to kaons, $\eta$-mesons etc, but as no reliable estimate exists in this regime~\cite{Clarke:2013aya}, our perturbative estimate is sufficient in this context.
For a given mass, the total width,  $\Gamma_{\phi}$, can now be obtained by summing over all the kinematically relevant decay modes. Analyzing the ratio $\Gamma_{\phi}/M_{\phi}$, we find that the relaxion is very narrow throughout the whole parameter space of our interest. For example,
\begin{equation}
 \frac{\Gamma_{\phi}}{\mphi } \simeq \left\lbrace 2\cdot 10^{-13},~10^{-5}\right\rbrace \cdot \sin^2\theta ~~\textrm{for}~~\mphi=\left\lbrace 0.1,~5\right\rbrace\gev \label{eq:GammaOverM}\,.
\end{equation}
For lighter masses, this ratio becomes even smaller. Hence, potential width effects do not arise. The ratio for intermediate masses between the two example values in Eq.~(\ref{eq:GammaOverM}) highly depend on the thresholds of those particles that the relaxion can decay into. The relaxion lifetime, $\tau_\phi=1/\Gamma_{\phi}$, is crucial in determining the applicability of various observational constraints. We show the lifetime as a function of $\mphi$ for different $\st$ values in figure~\ref{fig:ctau}.

 \bibliographystyle{JHEP}
 \bibliography{relaxion}

\end{document}